%% file: main.tex
\documentclass[twocolumn, astrosymb, twocolappendix,appendixfloats]{aastex701}

\input{preamble.tex}

\begin{document}

\correspondingauthor{Tai Withers}

\title{Ammonia Fitting of Multiple Velocity Components in High-Mass Star-Forming Regions}
\author{Tai Withers}
\email[show]{tai@withers.ca}
\affiliation{Department of Physics, Engineering Physics, and Astronomy, Queen's University, 64 Bader Lane, Kingston, ON, K7L 3N6, Canada}

\author{Sarah Sadavoy}
\email[]{sarah.sadavoy@queensu.ca}
\affiliation{Department of Physics, Engineering Physics, and Astronomy, Queen's University, 64 Bader Lane, Kingston, ON, K7L 3N6, Canada}

\author{Mike Chen}
\email[]{mkid.chen@gmail.com}
\affiliation{Department of Physics, Engineering Physics, and Astronomy, Queen's University, 64 Bader Lane, Kingston, ON, K7L 3N6, Canada}
\affiliation{NRC Herzberg Astronomy and Astrophysics, 5071 West Saanich Road, Victoria, BC V9E 2E7, Canada}

\author{James Di Francesco}
\email[]{James.DiFrancesco@nrc-cnrc.gc.ca}
\affiliation{NRC Herzberg Astronomy and Astrophysics, 5071 West Saanich Road, Victoria, BC V9E 2E7, Canada} \affiliation{Department of Physics and Astronomy, University of Victoria, Victoria, BC V8P 5C2, Canada}

\author{Rachel Friesen}
\email[]{rachel.friesen@utoronto.ca}
\affiliation{David A. Dunlap Department of Astronomy \& Astrophysics, University of Toronto, 50 St. George St, Toronto, ON M5S 3H4, Canada}

\author{Jared Keown}
\email[]{10jkeown@gmail.com}
\affiliation{Department of Physics and Astronomy, University of Victoria, Victoria, BC V8P 5C2, Canada}

\author{Erik Rosolowsky}
\email[]{rosolowsky@ualberta.ca}
\affiliation{Department of Physics, University of Alberta, 4-183 CCIS, Edmonton, Alberta, T6G 2E1, Canada}

\author{Helen Kirk}
\email[]{helen.kirk@nrc-cnrc.gc.ca}
\affiliation{NRC Herzberg Astronomy and Astrophysics, 5071 West Saanich Road, Victoria, BC V9E 2E7, Canada}
\affiliation{Department of Physics and Astronomy, University of Victoria, Victoria, BC V8P 5C2, Canada}

\author{Adam Ginsburg}
\email[]{adam.g.ginsburg@gmail.com}
\affiliation{Department of Astronomy, University of Florida, P.O. Box 112055, Gainesville, FL 32611, USA}

\author{Loren D. Anderson}
\email[]{loren.anderson@mail.uvw.edu}
\affiliation{Department of Physics and Astronomy, West Virginia University, Morgantown, WV 26506, USA}

\author{Paola Caselli}
\email[]{caselli@mpe.mpg.de}
\affiliation{Center for Astrochemical Studies, Max-Planck-Institut für Extraterrestrische Physik, Gießenbachstr. 1, 85748 Garching, Germany}

\author{Davide Elia}
\email[]{davide.elia@iaps.inaf.it}
\affiliation{INAF-IAPS, via del Fosso del Cavaliere 100, I-00133 Roma, Italy}

\author{Erin Figura}
\email[]{cfigura77@gmail.com}
\affiliation{Mathematics, Computer Science, and Physics, Wartburg College, Waverly, IA, USA}

\author{Hong-Li Liu}
\email[]{hongliliu2012@gmail.com}
\affiliation{School of Physics and Astronomy, Yunnan University, Kunming 650091, PR China}

\author{Anthony P. Marston}
\email[]{tmarston@sciops.esa.int}
\affiliation{ESAC/ESA, Camino Bajo del Castillo s/n, E-28691 Villanueva de la Ca\~{n}ada, Spain}
 
\author{Fr\'{e}d\'{e}rique Motte}
\email[]{frederique.motte@univ-grenoble-alpes.fr}
\affiliation{Univ. Grenoble Alpes, CNRS, IPAG, 38000 Grenoble, France}
 
\author{Stella Offner}
\email[]{soffner@utexas.edu}
\affiliation{Astronomy DepartSment, The University of Texas at Austin, 2515 Speedway, Stop C1400, Austin, TX 78712-1205, USA}
 
\author{Stefania Pezzuto}
\email[]{stefania.pezzuto@inaf.it}
\affiliation{INAF-IAPS, via del Fosso del Cavaliere 100, I-00133 Roma, Italy}
 
\author{Patricio Sanhueza}
\email[]{psanhueza@astron.s.u-tokyo.ac.jp}
\affiliation{Department of Astronomy, School of Science, The University of Tokyo, 7-3-1 Hongo, Bunkyo, Tokyo 113-0033, Japan}

\author{Eugenio Schisano}
\email[]{eugenio.schisano@inaf.it}
\affiliation{INAF-IAPS, via del Fosso del Cavaliere 100, I-00133 Roma, Italy}
 
\author{Nicola Schneider}
\email[]{nschneid@ph1.uni-koeln.de}
\affiliation{I. Physikalische Institut, University of Cologne, Zuelpicher Str. 77, 50937 Cologne, Germany}
 
\author{Ian W. Stephens}
\email[]{istephens@worcester.edu}
\affiliation{Department of Earth, Environment, and Physics, Worcester State University, Worcester, MA 01602, USA}

\begin{abstract}

We perform a multiple velocity component analysis of ammonia (\nhhh) gas from \acs{KEYSTONE}, a \acl{gbt} survey covering 11 \acfp{gmc} at distances from 0.7 to 2.7 kpc.
We extend existing codes to enable simultaneous fitting of up to three velocity components in \nhhh\ (1,1) with radiative transfer.
Using our multiple component fitting technique, we produce the best maps of the centroid velocity and velocity dispersion in these clouds to date.
Over the entire survey, multiple velocity components were seen in \percent{13} of pixels where we detect and fit \nhhh{} emission, primarily in clusters at the heart of the \acp{gmc}.
In these multiple component pixels, the median measured \nhhh{} linewidth decreased by a factor of 1.4 compared to equivalent single-component measurements, but there are also significant cloud-to-cloud variations.
Regions with multiple velocity components tend to be found in the densest zones of the \acp{gmc} and at the sites of high-mass star formation, indicative of multiple filaments, cores, or complex velocities associated with the formation of massive stars. 
Disentangling these velocity features will allow future quantitative analysis of coherent motions, such as infall and outflow, in these complex environments.
The multi-component data products are available on the CADC\footnote{\url{https://www.canfar.net/citation/landing?doi=26.0027}}.

\end{abstract}

\section{Introduction}\label{sec:introduction}

Stars form in dense cores within \acfp{gmc} comprised of gas and dust.
These small (typically defined as having radii \pc{\textless~0.1}) dense \citep[$n>10^5$ particles cm$^{-3}$,][]{coletta2025} cores are likely to produce either single or small groups of stars, although the correlation between core mass and final stellar mass(es) is unknown \citep{berginColdDarkClouds2007,offnerOriginUniversalityStellar2014}.
Cores are often found along or at the intersections of filaments --- elongated structures with lengths ranging from \pc{\textless~0.5} for nearby low-mass molecular clouds \citep{Jia2025} to \pc{10} for more distant higher mass clouds in the Galactic Plane \citep{hacarInitialConditionsStar2023}. 
Velocity gradients seen along some filaments indicate that mass may flow along them to `feed' cores \citep{schneider2010DR21,kirkFilamentaryAccretionFlows2013,perettoSDC13InfraredDark2014,yangDirectObservationalEvidence2023,PanSirong2024,friesenStabilityDenseCores2024}.
Dense cores can also cluster in $\sim 1$ pc clumps found at the intersection of filaments and hubs \citep{Xu2023}.
These clumps are key to current theories of cluster and high-mass star formation \citep{motteHighMassStarMassive2018, Beuther2025}.

Ammonia (\nhhh{}) line emission is a valuable tracer of cold, dense gas due to multiple transitions occurring at similar frequencies and its hyperfine structure \citep[][]{hoInterstellarAmmonia1983,berginColdDarkClouds2007}. 
The \ac{gas} used \nhhh{} emission to examine the gas motions in nearby star-forming regions such as the Taurus, Ophiuchus, Perseus, and Orion Molecular Clouds \citep{friesenGreenBankAmmonia2017,Pineda2025}. 
As these are some of the nearest clouds to the Sun (\pc{\textless~500}), these data were used to spatially resolve dense cores and clumps with in-depth virial analyses \citep{chenDropletsPressuredominatedCoherent2019,singhAreMassiveDense2021}. 
In a re-examination of the initial \ac{gas} data, \citet{chenVelocityCoherentFilamentsNGC2020} fitted multiple velocity components to the \nhhh{} (1,1) data of NGC 1333 in Perseus (see \refsec{sec:MCFitting} for methodological details). 
Using their updated parameter maps, \citet{chenVelocityCoherentFilamentsNGC2020} identified ten filaments in position-position-velocity space, and analyzed gradients in velocity along them. These additional components were found in the innermost regions with the brightest emission, and the ability to extract them improved the identification and analysis of kinematically coherent structures.
Indeed, these filaments are consistent with current theories in which gas is accreted onto filaments and then transported to clumps and cores \citep{chenostriker2014,RSmith2016,clarke2017,chenVelocityCoherentFilamentsNGC2020}.

To facilitate comparisons between high- and low-mass star-forming regions, the \keystone{} survey \citep[][hereafter \citetalias{keownKFPAExaminationsYoung2019}]{keownKFPAExaminationsYoung2019}, conducted a similar \nhhh\ survey as GAS but for 11 high-mass GMCs, including high overlap with the \ac{hobys} survey \citep{motteInitialHighlightsHOBYS2010,motteHighMassStarMassive2018}, complementing the existing dust observations with gas tracers.
The first \keystone{} data release included 835 dense cores and clumps across the \acp{gmc}.
These objects were identified using peaks in \nhhh{} emission, and \citetalias{keownKFPAExaminationsYoung2019} find that \percent{$\approx60$} of them are gravitationally bound. 

Nevertheless, the initial analysis of the \ac{KEYSTONE} survey was based on single velocity component fits to the \nhhh\ (1,1) data. 
Since high-mass star-forming regions are more turbulent and complex than their lower-mass counterparts, we expect that \nhhh\ emission in \acp{gmc} will also have multiple velocity components similar to what \citet{chenVelocityCoherentFilamentsNGC2020} found in low-mass star-forming regions. 
Indeed, other studies have identified multiple velocity components in \acp{gmc} with high-density tracers \citep[e.g.,][]{sokolovMulticomponentKinematicsMassive2019,choudhuryTransitionCoherentCores2021,chenFilamentAccretionFragmentation2024}, but most of these works are limited case studies of individual regions of interest, and survey-scale studies have been mostly limited to single-component analysis with few studies applying multiple components \citep[e.g.,][]{Hogge2018}. 
The potential insights obtained from fitting multiple components motivates a re-analysis of the \keystone{} data. 

Here, we present a new in-depth analysis of the \keystone{} data using multiple-component fitting of the \nhhh\ data, building on the work done by \citetalias{keownKFPAExaminationsYoung2019} and expanding on the framework and algorithms of \citet{chenVelocityCoherentFilamentsNGC2020}.  
This paper is structured as follows:  In \refsec{sec:observations}, we summarize the \keystone{} observations used in this analysis.  
In \refsec{sec:methods}, we detail the methods used for fitting up to three velocity components and explain our algorithm for selecting the best models.
The results of our fitting are given in \refsec{sec:results}, along with the details of our data products.
\refsec{sec:discussion} discusses our results in a wider context, and finally, in \refsec{sec:summary}, we present our conclusions. 

\section{Observations}\label{sec:observations}

We use \nhhh{} (1,1) observations taken by the \keystone{} survey from October 2016 through March 2019.
These data and an initial analysis were published by \kt.
The observations used the \ac{kfpa} instrument on the \ac{gbt} and totalled 356.25 hours on sky over 11 \acp{gmc}. 
\reftab{tab:GMC-Summary} gives the map centre, map area, and distance to each of the regions in the survey.

\begin{table*}[htb!]
    {\centering
    \caption{\acs{KEYSTONE} Target Clouds.\label{tab:GMC-Summary}}
\input{Tables-TargetClouds}}
\tablecomments{
    References used for distance: (1) \citet{dharmawardenaThreedimensionalStructureGalactic2023}, (2) \citet{zuckerCompendiumDistancesMolecular2020}, (3) \citet{movsessianNewHerbigHaro2020}, (4) \citet{moscadelliTrigonometricParallaxesMassive2009}, (5) \citet{muzicStellarPopulationRosette2022}, (6) \citet{chibuezeAstrometryH2OMasers2020}.
    Values of rms are medians over the cloud using the spectrally smoothed (\kms{0.14}) data cubes and calculated based on our fitting routine products.
    \textsuperscript{a} W3 is split into W3 West, which contains the HII region KR140, and W3 Main.
    }
\end{table*}

The maps were observed with frequency switching over $10\arcmin \times 10\arcmin$ tiles.
The Moon was used for flux density calibration.
The data reduction was performed with \formatSoftware{gbtpipe}\footnote{\url{https://github.com/GBTSpectroscopy/gbtpipe}}, and the observations were gridded to
$8.8\arcsec$ pixels, which oversamples the $32\arcsec$ beam compared to Nyquist sampling \citepalias{keownKFPAExaminationsYoung2019}.
Spectra within a beam are correlated, with one beam covering approximately 4 pixels across.
The spectrum of each pixel was calculated with a weighted average of all seven \ac{kfpa} beams following the pipeline created for \ac{gas} \citep{friesenGreenBankAmmonia2017}.
Further details on the observations and data reduction pipeline can be found in \kt.
The final spectral resolution of the cubes is \hz[k]{5.7} or \kms{0.07}, covering a bandwidth up to 7.85 MHz (\kt{}) around the \nhhh{} (1,1) central frequency of 23.694 GHz \citep{lovasNISTRecommendedRest2004}.
These data cubes, along with the results of the fitting done by \kt{} are publicly available.
We also use the single-component \nhhh{} line fitting done by \kt{} to help constrain our multi-component fitting process.

For our analysis, the data cubes are spectrally convolved by a factor of two and resampled along the spectral axis to reduce the impact of noise and provide smoother spectra for fitting.
The resulting spectral resolution is \hz[k]{11.4} or \kms{0.14}.
Several of the clouds were also broken into smaller regions (called subclouds or subregions) to improve visualization and fitting.
A complete list of the modified clouds and the pixel coordinates used to define map boundaries can be found in \refapp{app:subclouds}.
The coordinates for these subclouds were chosen based on \momzero{} maps, with boundaries intended to maximize the fraction of pixels that contained emission. 
\reffig{fig:moment_zero_with_overlays} shows an example \momzero{} map from \kt{} for Cygnus X North with the three subregions labeled in pink as A, B, and C.
The smaller grey boxes indicate the sections of each subregion with no detectable emission based on the \momzero{} maps that we use to estimate noise for each the subregion.
Similar maps for the other regions can be found in \refapp{app:subclouds}.
\input{Figures-MomentZeroMaps_Example}

\section{Methods}\label{sec:methods}

This Section details our algorithm for fitting and selecting up to three \nhhh{} velocity components, including how our method builds on and expands from previous work. 
Unlike previous multi-component fits \citep{Hogge2018}, we perform multi-component fitting with radiative transfer such that optically thick foreground emission can affect the line profiles of background emission.
Broadly, \nhhh{} data are fitted based on four parameters: centroid velocity $\vlsr$, velocity dispersion $\vdobs$, excitation temperature $\tex$, and optical depth $\tau$. 
The observed intensity, as measured by the main beam temperature $\tmb$, is calculated as a function of frequency ($\nu$) with \refeq{eq:TMB(nu)1c} \citep{friesenGreenBankAmmonia2017} for a single-component spectrum under \ac{lte}
\begin{equation}\label{eq:TMB(nu)1c}
    \tmb(\nu) = \Big[ J_\nu (\tex) - J_\nu (\tcmb) \Big] \Big[ 1 - \exp{\left(-\tau(\nu)\right)}\Big],
\end{equation}
where $\tcmb = \K{2.73}$ is the \ac{cmb} temperature and
$J_\nu$ is the Rayleigh-Jeans equivalent temperature defined as, 
\begin{equation}\label{eq:Jnu(T)}
    J_\nu(T) = \frac{h\nu/\kb}{\exp{(h\nu/\kb T)}-1},
\end{equation}
where $h$ is the Planck constant and $k_B$ is the Boltzmann constant.
Following \citet{ginsburgPySpecKitPythonSpectroscopic2022}, we define the optical depth profile as a sum over each of the 18 hyperfine components,
\begin{equation}
    \tau(\nu) = \sum_{i=1}^{18} \tau_{\oneone} W_i \phi_i(\nu),
\end{equation}
where $W_i$ is an associated weighting factor  \citep[tabulated in][]{kukolichMeasurementAmmoniaHyperfine1967}  and the line profile of each hyperfine component $\phi_i(\nu)$ is a Gaussian centered on each hyperfine line with a width given by $\vdobs$. 

The total optical depth of the spectrum $\tau_{\oneone}$ is given by \citet{ginsburgPySpecKitPythonSpectroscopic2022} as 
\begin{equation}
    \tau_{\oneone} = P \frac{c^3 A_{(1,1)}}{8\pi \nu_0^3\sqrt{2\pi}} \frac{1-\exp{(h\nu/\kb \tex)}}{1+\exp{(h\nu/\kb \tex)}},
\end{equation}
where $A_{(1,1)}$ is the Einstein A coefficient of the transition, $c$ is the speed of light, $\nu_0$ is the rest frequency of the transition, and $P$ is the population in the excited state which is proportional to the \nhhh{} column density and the transition partition function.
Since the spectrum is made up of multiple hyperfine components, the \nhhh{} transition is not considered `optically thick' until $\tau_{\oneone} \gtrsim 4$ \citep{rosolowskyAmmoniaSpectralAtlas2008}. 
For brevity, we use $\tau$ as a shorthand for $\tau_{\oneone}$, and it is this $\tau$ that is used in generating a model.
For all of the \nhhh{} model fitting performed, we employ the \pyspeckit{} fitter, which performs Levenberg-Marquardt least-squares minimization and reports parameter errors as the square root of the diagonal of the covariance matrix \citep{ginsburgPySpecKitPythonSpectroscopic2022}.

\subsection{Multicomponent Fitting}\label{sec:MCFitting}

The previous KEYSTONE data release (\kt{}) assumed that the \nhhh{} spectra could be fit by a single velocity component, which means that the emission comes from one homogeneous, isothermal packet of gas moving at a uniform speed along each line of sight \citep[e.g.,][]{friesenInitialConditionsClustered2009,pinedaDirectObservationSharp2010,bihrKinematicThermalStructure2015,friesenGreenBankAmmonia2017}.
Nevertheless, \acp{gmc} are complex environments, often containing networks of filaments connecting into hubs as well as feedback from clusters and massive stars \citep[e.g.,][]{motteHighMassStarMassive2018, pinedaBubblesFilamentsCores2023}. 
These factors complicate the gas motions such that a detailed look should include multiple velocity components.
Indeed, observations have shown that even less massive clouds can commonly feature emissions that cannot be adequately captured by single velocity models \citep[e.g.,][]{chenFilamentAccretionFragmentation2024}.

\citet{chenVelocityCoherentFilamentsNGC2020} developed the \mufasa{} routine to fit \nhhh{} emission with up to two velocity components with radiative transfer through each component.
In this approach, the emission is modeled as independent slabs, with the background emission moving sequentially through the slabs towards the observer, one slab at a time. 
Thus we revise \refeq{eq:TMB(nu)1c} according to \citet{chenVelocityCoherentFilamentsNGC2020}, where a spectrum is generated for $N$ velocity components by evaluating $I_1, I_2, ..., I_N$:
\begin{gather}
    I_i = J_\nu(T_i) \Big( 1 - e^{-\tau_i(\nu)} \Big) +I_{i-1}e^{-\tau_i(\nu)}\\
    \tmb(\nu) = I_N - J_\nu (\tcmb)
\end{gather}
where $I_0 = J_\nu(\tcmb)$. 
With this approach, \mufasa{} is able to account for instances of an optically thick foreground component absorbing emission from a background gas packet, as discussed in \refsec{sec:ScarPlusConsiderations}.
After completing both a one- and two-component fit, \mufasa{} uses a statistical test to determine what number of components best describes the observed line emission at each location. 

\subsection{Three-Component Fitting}

\begin{figure*}[htb!]
    \centering
    \includegraphics[width=0.9\linewidth]{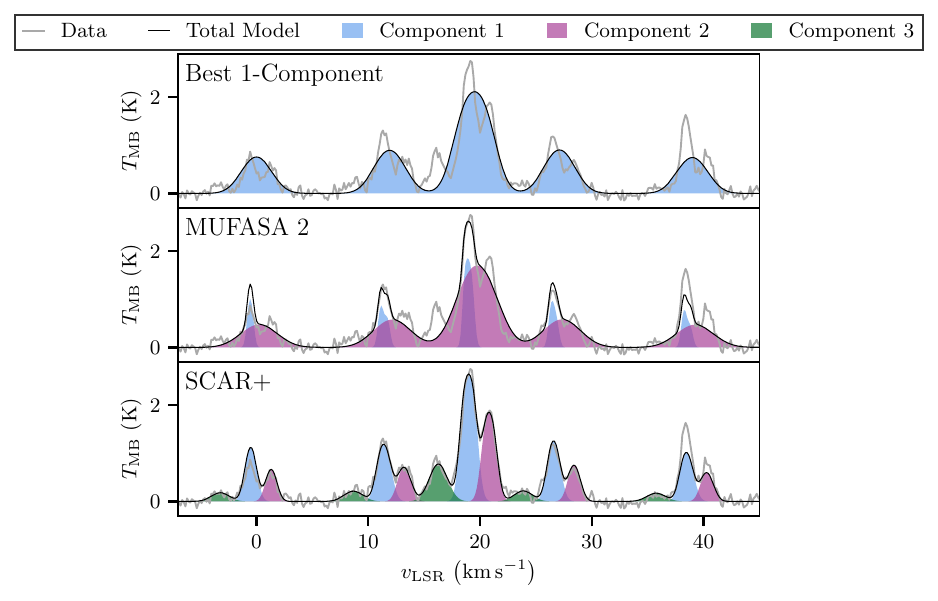}
    \caption{\label{fig:RequiresThirdComp}Example spectrum in M17 showing the necessity of three velocity components. The upper panel shows a single-component fit, the middle panel shows the best-fit two velocity component model with \mufasa, and the lower panel shows the best-fit three-component fit from our \scarplus{} routine. 
    Notably the component widths change significantly between the models, exemplifying the need for a three-component fit in order to characterize the gas kinematics.
    This pixel also shows evidence of \acl{ahe}, as indicated by the mismatch in the heights of the satellite groups between the spectrum and models.
    We discuss this phenomenon further in 
    \refsec{sec:ahe}.
    }
\end{figure*}

When applying \mufasa{} to the \keystone{} clouds, we found that some of the more complex regions were poorly modeled by two (or less) velocity structures.
\reffig{fig:RequiresThirdComp} shows an example spectrum and model fit for one of these complex regions, where we can see that neither one-component nor two-component models fully capture the data.
As such, we developed an extension to \mufasa{} called the \scarplus{} to fit additional velocity structures.  Broadly, we developed \ac{SCAR} as an iterative extension to \mufasa{} that enables us to fit up to three velocity components simultaneously.

\reffig{fig:Flowchart3cFitting} outlines the primary stages of \scarplus.  Like \mufasa, \scarplus{} performs multiple different fits to each spectrum, e.g., one, two, and three components, and then utilizes statistical testing to determine which fit is best.  
In brief, the first set of steps are designed to perform an initial fit of one and two components from which the data can be prepared for a three component fit.  
The second set of steps (within the grey box) perform the three-component fitting.  
The three-component fitting is done on smaller subregions and as such, this set of steps must be iterated over the entire map.

In the following subsections, we describe the \scarplus{} implementation.
\refchap{sec:TwoComponentFitting} discusses Steps 0-3, \refchap{sec:ThreeComponentFitting} details Steps 4-8, 
\refchap{sec:ScarPlusConsiderations} explains some additional considerations for the use and application of \scarplus,
and finally \refchap{sec:statistical_testing} explains the statistical tests used to select the best-fit model.

\begin{figure}[htb!]
    \centering
    \input{Figures-Flowchart3cFitting}
    \caption{\label{fig:Flowchart3cFitting} Depiction of the \scarplus{} process to generate one-, two-, and three-component parameter cubes. 
    The grey box highlights the processes which are repeated over each small region, and make up our novel \scarplus{} pipeline.  
    This flowchart ends with the completion of up to three-component fits. 
    Statistical testing to determine the best-fit model is performed separately and discussed in  \refsec{sec:statistical_testing}.}
\end{figure}

\subsubsection{Data Preparation and Two-Component Fitting} \label{sec:TwoComponentFitting}

Step 0\footnote{We label the first step ``zero'' as this action may not be necessary depending on the dataset. } 
of the process is an \ac{ecr} where we perform a single-component fit to the data using the values from \kt{} as initial guesses. We require this re-fit because we are using spectrally-smoothed data compared to \kt{} and a small fraction ($\approx 0.7\%$) of the spectral line fits in \kt{} had unusually high error values (by \textgreater{} factor of 10, particularly in the velocity dispersion), seemingly unrelated to the quality of the fit. 
The cause of the large error values in \kt{} is unknown, but they affect our ability to use those data as constraints in the \scarplus{} pipeline and to consider the original single-component fits as best-fit models in our analysis. 
The resulting \ac{ecr} models generally showed good agreement\footnote{\percent{95} of re-fitted centroid velocity values were within \percent{2} of those originally reported.} with \kt{} such that all future comparisons to the parameters reported by \kt{} are comparisons to the \ac{ecr} results unless otherwise noted.

For Step 1, the data are masked to reduce computation time spent attempting to fit spectral lines where none are present.
First, we blanked noisy pixels, defined as being within three pixels of a map edge or abnormally bright over many channels.
We assume that the brightest \nhhh{} spectra will only have high intensities in a few channels.
Thus any pixel which is identified as having the highest intensity over the entire map in at least 5\% of the channels ($\approx$ 30) is masked out.
Second, we masked out regions consistent with little to no emission. 
For this masking, we used the \momzero{} maps generated by \citetalias{keownKFPAExaminationsYoung2019}, which were measured over spectral channels that were indicated by their line fitting to have \nhhh{} emission.
These \momzero{} maps exclude channels which contain emission from an un-fitted component, but this does not seem to impact our masking.  Using these maps, we masked out pixels with \ac{snr} \textless~3, where the noise was defined in the regions given in \refapp{app:subclouds}.
Finally to clean the mask, isolated pixels were removed and single-pixel holes in the masks were filled in, such that these spectra would be included in the fit.

Following the above masking process, Step 2 uses the \mufasa{} package to fit the \nhhh{} spectra.
\mufasa{} was built for use on data from \ac{gas}, which primarily observed quiescent low-mass star-forming regions.
As the \keystone{} survey targeted high-mass star-forming clouds, these regions have different gas dynamics.
As a result, we imposed a different set of parameter limits for the \keystone{} clouds compared to the nominal values used by \mufasa{}.
\reftab{tab:DefaultParameterLimits} compares our parameter limits to the values used internally in \mufasa.
As \mufasa{} has hard-coded parameter limits, we could not adjust these values at the fitting stage and instead confined the output parameters.
The fitting results were checked for values which lay outside of our imposed limits, and in these cases, the parameters were set to the relevant maximum or minimum value.
The minimum velocity dispersion is set to be the spectral channel width, while the maximum is chosen at \kms{1.75} to reflect that turbulence exceeding \kms{1.5} is not expected in high-mass star-forming regions.
The excitation temperature is capped at \K{50}, which is slightly more permissive than the default maximum of \K{40} used by \mufasa{} and the maximum optical depth is significantly more restricted than the default of 30. 
At optical depth values greater than a few, the spectra are considered optically thick regardless of the exact value.
Reducing the maximum optical depth allows for optically thick solution while avoiding badly-converged fits.

These limits were chosen to reflect more accurately the high-mass star-forming environments in which they were applied or to limit parameter space to ease fitting, particularly since the incorporation of radiative transfer can, in some cases, bias towards unphysical results. 

\begin{table}[htb!]{\centering    \caption{\label{tab:DefaultParameterLimits}Parameter Limits}
    \begin{tabular}{lll} \toprule
             & \th{Minimum}      & \th{Maximum}   \\ \midrule
$\vlsr$ (\mbox{km s$^{-1}$})   & None         & None      \\
$\vdobs$ (\mbox{km s$^{-1}$}) & 0.14 (0.07)  & 1.75 (2.5)\\
$\tex$ (K)               & 3 (3)        & 50 (40)   \\
$\tau$                            & 0.1 (0.1)    & 7.5 (30)  \\
    \bottomrule \end{tabular}}
\tablecomments{Numbers in parentheses indicate the nominal \mufasa{} values.}
\end{table}

In addition to placing limits on the parameters output by \mufasa, for the multi-component fits, we consider only those which have at least one component with \ac{snr} $\geq 7$. 
For the multi-component fits, we additionally reject all components if they all  have \ac{snr}~\textless~7.
This rejection ensures that multiple components are only selected for bright spectra, without rejecting components that have a lower peak but nonetheless contribute to the overall shape of the model via broad linewidths or higher optical depths.
We then repeat the \ac{snr} cut performed on the \ac{ecr} by rejecting any individual components with a line peak \ac{snr}~\textless~3.
In this case we measure \ac{snr} as the maximum brightness temperature of the model divided by the rms value of the pixel.
Using this method, the rms is model dependent (e.g., the channels used to measure the rms depend on the model, as the model and its parameters determine the on-line channels). 
We then measure the rms as the standard deviation following 
\begin{equation}
\text{rms} \approx 1.4826 \times \text{}\left(\abs{X_i - \Tilde{X}}\right)
\end{equation}
where $\Tilde{X}$ is the median of the set of values $X_i$, which corresponds to line-free channels (e.g., the spectral channels that have a model brightness $< 0.0125$ K). 

Our final processing stage, Step 3, generates a new fitting mask containing only pixels for which \mufasa{} generated a model.
This mask is a subset of the mask from Step 1 (containing \percent{$\approx 70$} of the Step 1 unmasked pixels), and supersedes that initial mask for all future steps. 
The presence of a \mufasa{} fit appears to be a good litmus test for whether or not emission is potentially detectable using our routine.

\subsubsection{\acs{SCAR}} \label{sec:ThreeComponentFitting}

In this subsection, we describe the \scarplus{} process itself (Steps 4 -- 8 in \reffig{fig:Flowchart3cFitting}). 
We use \scarplus{} to fit up to three velocity structures in the \keystone{} clouds. 
Broadly, the pipeline operates by first removing a single, bright velocity component from all pixels in a subgrid to reduce the maximum number of components from three to two so that \mufasa{} can then run normally.
Thus, the spectral components are identified in batches which are then stitched together to perform one final multi-component fit. 
We note that this approach could be further modified to fit four or more velocity components, but we did not find cases where more than three components were necessary in the \keystone{} clouds.\footnote{Based on visual inspection of both the fitted models and the residuals, we found no cases where four or more components appeared to be necessary.}

Throughout \mufasa{} and \scarplus{}, we use the Levenberg-Marquardt method for least-square fits, which requires good initial guesses to find the global minima effectively. \mufasa{} identifies good initial guesses in a multi-step process.
First, the spectral cube data are spatially convolved with a Gaussian to smooth them by a factor of two. 
Second, a single-component fit is made with the smoothed cube  and the results from this fit are used as initial guesses for fitting the data at full resolution.
This process is done separately for both one- and two-component fits. 
For the two-component fits,
\mufasa{} generates initial guesses by copying the guesses for the one-component fit and spreading the values apart, using one guess slightly higher than the one-component fit result, and one guess slightly lower.
This method of copying and spreading is ineffective for more than two components as the fitting routine is generally unable to iterate from these guesses to a viable result.
Therefore it was necessary for us to develop a new routine that also produced realistic initial guesses for up to three components, which we outline at the appropriate stages.

\begin{figure}[htb!]
    \centering
    \includegraphics[width=\linewidth]{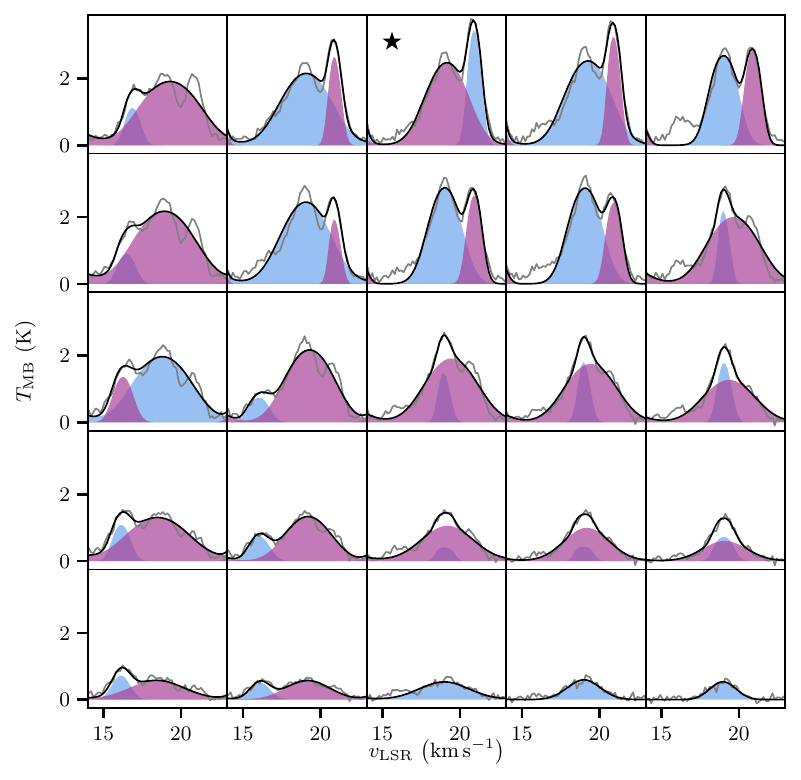}
    \caption{\label{fig:SubgridStep5}
    A \scarplus{} subgrid of pixels from M17, along with the best-fit results from \mufasa{} (Step 2).
    Only the central hyperfine group is shown.
    The thin blue component in the starred pixel is the component identified in Step 5 as being the highest-intensity within the subgrid, which will be used to define the parameter space for the Step 6 single-component fitting of all pixels in the grid. 
    Note that the colour of a component has no physical meaning, and is used only to distinguish the two. 
    }
\end{figure}

Our pipeline begins (Step 4) with gridding the data into 5$\times$5 pixel subgrids, such that  the subgrids are slightly larger than the \ac{gbt} beam. 
Since the spectra within a beam are correlated, we assume that all spectra across the subgrids will have similar spectral features and therefore, we can safely assume that a single bright component in one pixel should be present in all pixels of the subgrid.  
\reffig{fig:SubgridStep5} shows an example subgrid along with the \mufasa{} results from Step 2. 
In \reffig{fig:SubgridStep5} and the following figures, we show only the central hyperfine group for clarity, but throughout the algorithm all hyperfine lines are included in the fit.

After Step 4 and before proceeding with the rest of the \scarplus{} pipeline, the subgrid is inspected. 
We require that at least 10 pixels in the subgrid have best-fit models from the \mufasa-only stage (Steps 1-2), and that the brightest individual component in the best-fit models has a peak of $\geq 0.5$~K. 
This requirement ensures that there are sufficient pixels to identify and measure the brightest component.

For Step 5, we use the prior best-fit \mufasa{} model to identify the brightest velocity component (by peak intensity) over the entire subgrid. For example, the blue component in the starred panel in \reffig{fig:SubgridStep5} is the brightest component of that subgrid.
If the best-fit model at a pixel has two components, we consider these components separately and take the brighter of the two, as is the case for the pixels in \reffig{fig:SubgridStep5}. 

\begin{figure}[htb!]
    \centering
    \includegraphics[width=\linewidth]{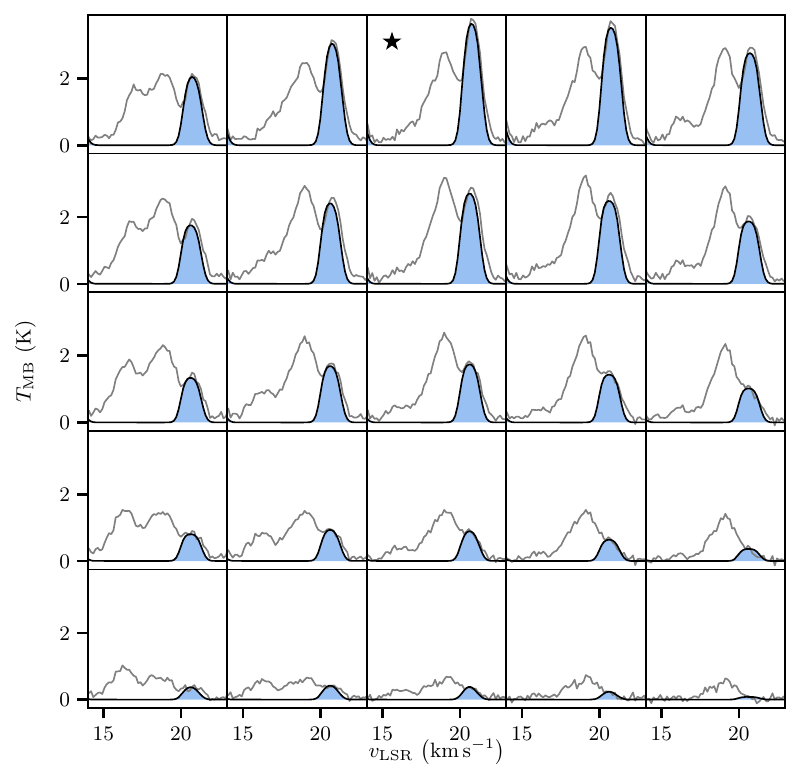}
    \caption{\label{fig:SubgridStep6}
    Outcome of Step 6. 
    Spectra are the same as \reffig{fig:SubgridStep5}, but the blue models are fitted using a restrictive parameter space generated from the brightest component (shown by the star symbol) identified in Step 5.
    These blue models are subtracted from the data to create a residual cube.
    }
\end{figure}

In Step 6, we fit only the identified `brightest component' to every pixel in the subgrid using very restrictive parameter limits to ensure that only this component is fitted.
\reffig{fig:SubgridStep6} shows the result of this fitting applied to the same grid of pixels from \reffig{fig:SubgridStep5}.  To ensure only the `brightest component' is fitted, we use the best-fit values from the original \mufasa{} model as the initial guesses for all pixels in the subgrid with strict restrictions on the velocity centroid and velocity dispersion ranges.
Since the subgrids are slightly more than a beam-width, we can assume that the velocity components within the subgrids should be correlated such that the line centroids and widths should not change drastically over the entire subgrid.
We therefore use parameter limits of \kms{$\vlsr \pm$~0.28} (two spectral channels) and \kms{$ \vdobs \pm$~0.025}.  These tight parameter limits ensure that only the brightest component is fitted as a contiguous structure and that any other velocity components are excluded.
For pixels where the best-fit `brightest component' velocity dispersion was greater than \kms{1}, we reduced it to \kms{0.75} to increase the likelihood that the \ac{SCAR} method would separate multiple blended components rather than a fit a single wide component. After performing this restricted single-component fit, we create a residual cube of this subgrid by subtracting the fitted bright component from each pixel.

\begin{figure}[htb!]
    \centering
    \includegraphics[width=\linewidth]{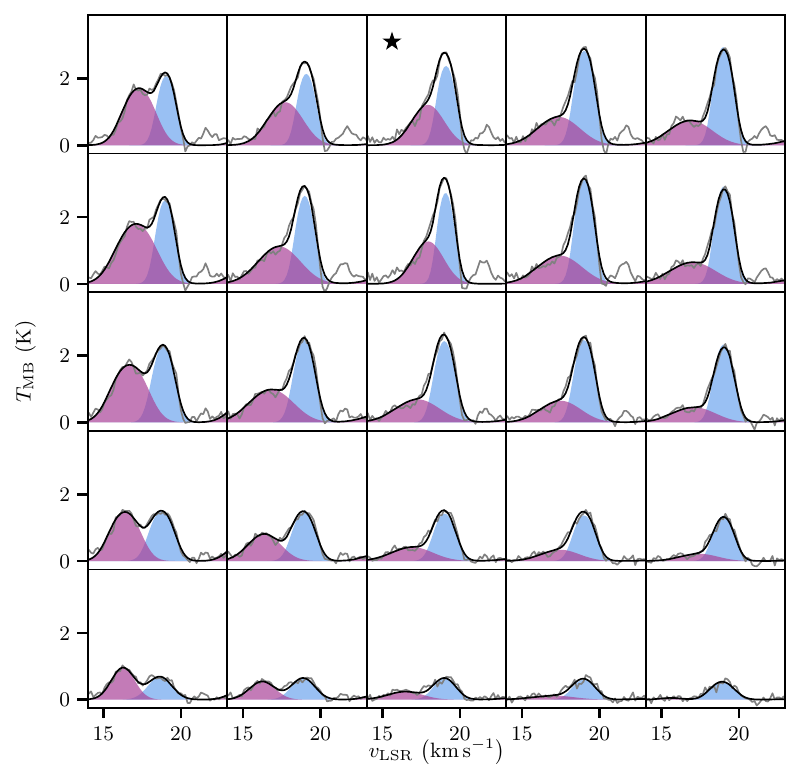}
    \caption{\label{fig:SubgridStep7}
    Outcome of Step 7. 
    Spectra are the residuals from the fit performed in Step 6, and models are generated by fitting these residual spectra with \mufasa.
    For these fits, the initial guesses for the models are generated by \mufasa{} using moment maps of the residual cube itself.
    }
\end{figure}

In Step 7, we fit the residual cubes to recover the remaining spectral lines.  The spectra in the residual cube should now contain zero, one, or two components as opposed to the one, two, or three components present in the original data, because the `brightest component' has been removed from each pixel.
Since the maximum number of components is two, we can apply \mufasa{} normally, using the parameter limits outlined in \refchap{sec:TwoComponentFitting}.
\reffig{fig:SubgridStep7} shows the results of Step 7, where all remaining emission in the residual cube is captured using \mufasa. 
Note that the brightest component (shaded region in \reffig{fig:SubgridStep6}) has already been subtracted from the spectra in \reffig{fig:SubgridStep7}.

After Step 7, we have two sets of parameters that are specific to the subgrids: the results of the brightest single-component fit and the results of the \mufasa{} fit to the residuals.
For Step 8, we combine these parameters by fitting all three components simultaneously, again with strict limits on the centroid velocity and velocity dispersion.
At this stage, we expect minimal changes in the kinematic parameters, and this fitting is done to account for the radiative transfer between the components.  For example, \reffig{fig:SubgridStep7} shows excess unfitted emission around the \kms{22} range of several pixels, showing that the brightest-component fitting of Step 6 was imperfect.  The purpose of Step 8 is to improve the fit using all three components and radiative transfer at once.  Since \pyspeckit{} can only accept a single set of limits for the parameters, we must define the kinematic parameter limits to encompass all possible velocity components.
We describe how these parameter limits are derived in \refapp{app:trifitparlims}.
The excitation temperature and optical depth limits are free to vary within the restrictions given by \reftab{tab:DefaultParameterLimits}.

\begin{figure}[htb!]
    \centering
    \includegraphics[width=\linewidth]{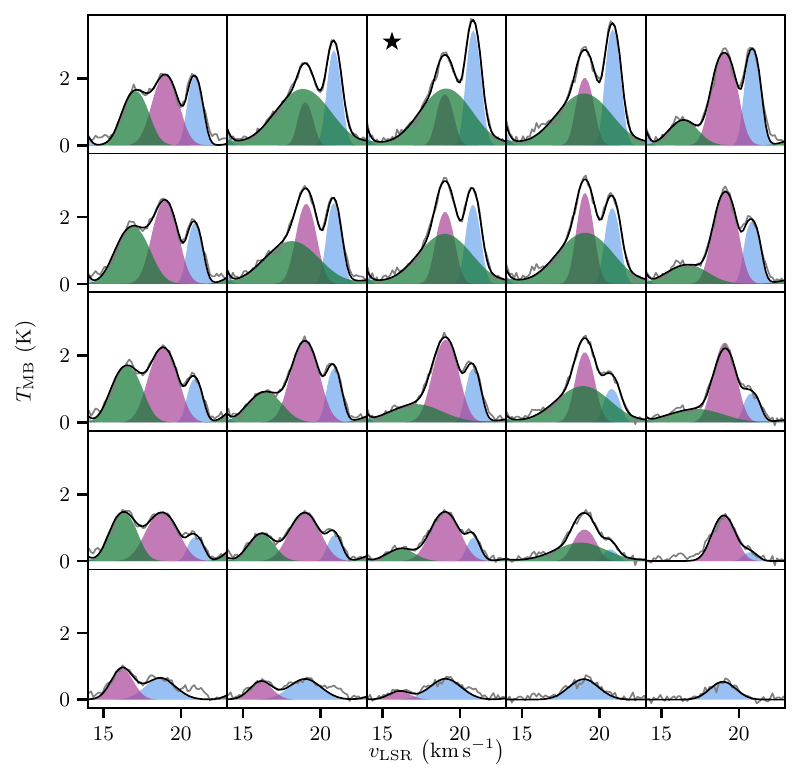}
    \caption{\label{fig:SubgridStep8}
    Outcome of Step 8.
    The models from Step 6 (\reffig{fig:SubgridStep6}) and Step 7 (\reffig{fig:SubgridStep7}) are used to build a three-component fit in a restricted parameter space.
    Colour is used solely to indicate the different components, and component colour has no continuity between the previous figures and this one.
    }
\end{figure}

\reffig{fig:SubgridStep8} shows the results of the three-component fit for our example subgrid. Following \mufasa{}, we apply both our \ac{snr} cut, where multi-component pixels which have only dim components (\ac{snr}~\textless~7) are removed, as well as our
componentwise \ac{snr} $< 3$ cut, which results in some \scarplus{} fits having one or two components, as the others were removed. 
As with the application of these \ac{snr} cuts to the \mufasa{} output, we expect that some fraction of present, but not well-detected, multi-component emission will be eliminated as a result of these criteria. 

Nevertheless, we note that even our sample subgrid in  \reffig{fig:SubgridStep8} shows pixels with bright but poorly constrained or unnecessary third components. We use a final reduced component fit and statistical tests to remove these cases (see Sections \ref{sec:ReducedFitting} and \ref{sec:statistical_testing}). 

Finally, at the end of the pipeline, we have \scarplus{} fits for individual subgrids.  In the final step the parameter cubes from each of the subgrids are stitched together to re-form the full-size parameter maps. Our complete pipeline for \ac{SCAR} is publicly available as \formatSoftware{ScarPlus}\footnote{\url{https://zenodo.org/records/23487470}} under a GNU General Public License version 3, and the version used for this work will be archived.

\subsection{\acs{SCAR} Considerations}\label{sec:ScarPlusConsiderations}

\subsubsection{Note on Component Ordering}\label{sec:ComponentOrdering}

\begin{figure*}[htb!]
    \centering
    \includegraphics[]{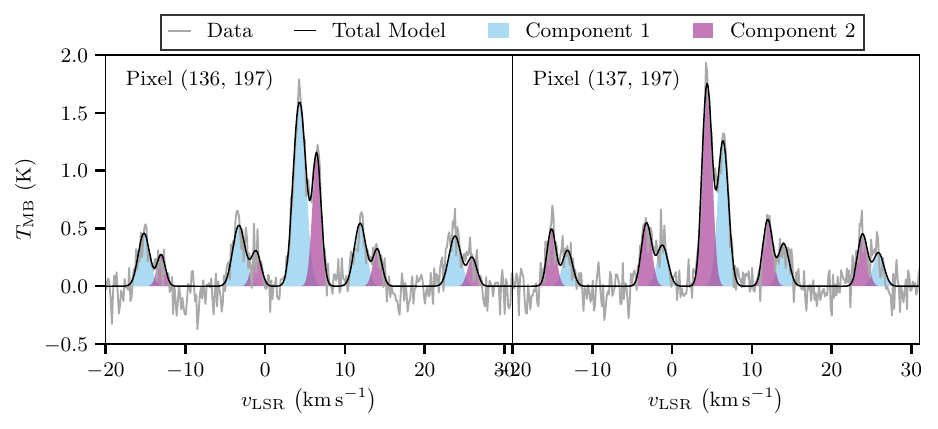}
    \caption{\label{fig:ComponentSwitchingExample}
    Spectra and \mufasa{} fits to two pixels in NGC2264, in which the component labelling (indicated by chosen colour) has swapped.
    The work presented in this paper makes no attempt to apply coherent ordering of components.
    }
\end{figure*}

While the inclusion of radiative transfer enforces some element of component ordering, \mufasa{} does not order its components in any particular way on a multi-pixel scale.
As an example of this, \reffig{fig:ComponentSwitchingExample} shows spectra and fits for two adjacent pixels in MonR1.
Both pixels contain velocity components at \kms{$\approx$~4} and \kms{$\approx$~6}, but the labeling in the code of which component is which swaps.
The primary result of this ambiguity is that property maps are prone to sharp jumps as the component labels swap arbitrarily.

\begin{figure*}[hbt!]
    \centering
    \includegraphics[]{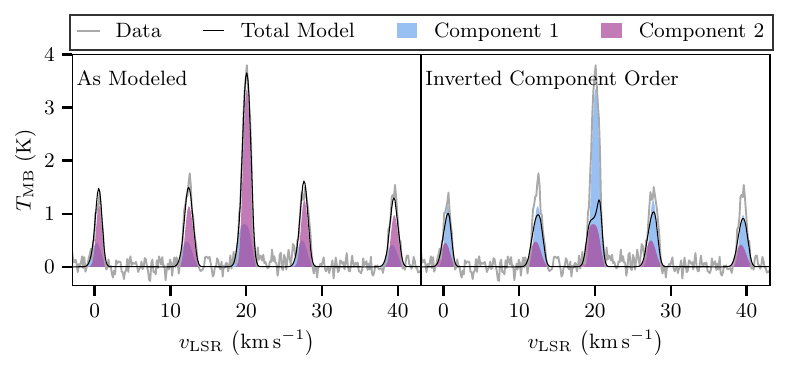}
    \caption{\label{fig:RadiativeTransfer} The effect of component ordering on the radiative transfer models for an example spectrum in M16. \textit{Left:} a two-component fit as modeled by our pipeline. 
    \textit{Right:} the same modeled components, but with the order of the components reversed.  This inverted component is not a fit to the observations but is meant to show the affects of radiative transfer on component order.
    }
\end{figure*}

While in some cases the component ordering has little to no effect on the resulting model (e.g., with non-overlapping spectra), this is not always the case.  {In particular, component order matters significantly if one or more components is optically thick.} 
Since most of the KEYSTONE clouds show optically thin \nhhh{} emission for all components, component ordering has little to no effect on the resulting model.  Nevertheless, roughly \percent{13.5} of the pixels have optically thick solutions ($\tau > 3.5$), and in those cases, component order can matter significantly. 
For example, \reffig{fig:RadiativeTransfer} shows how a different component order can affect the resulting model spectra for the same pixel in M16. 
The left panel shows the nominal spectrum from the \scarplus{} pipeline, whereas the right panel shows the same fitted components, but with their order reversed. 
The inverted case is not meant to be a fit to the observations, but is meant to highlight how the component order in a pixel can affect the produced spectrum, e.g., when a brighter component is blocked by an optically thick fainter component.  Thus, while component order does not matter from one pixel to the next in most cases (e.g., \reffig{fig:ComponentSwitchingExample}), it is still important to maintain a consistent order when generating radiative transfer models.

\subsubsection{Additional Reduced-Component Model}\label{sec:ReducedFitting}

\begin{figure*}[htb!]
    \centering
    \includegraphics[width=0.49\linewidth]{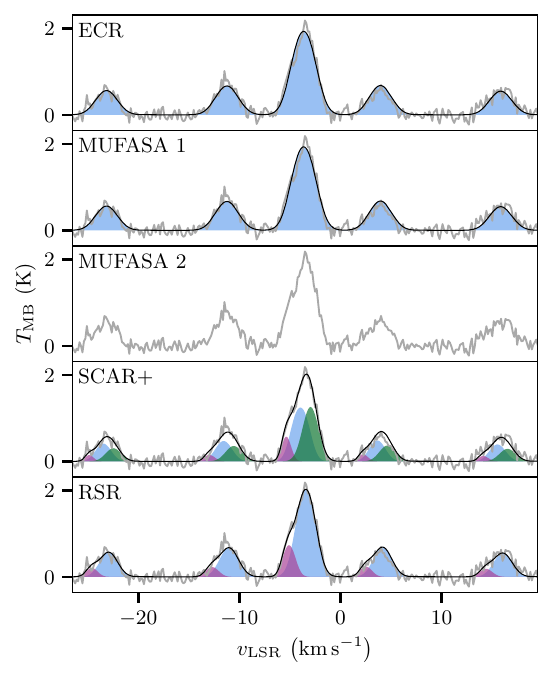}
    \includegraphics[width=0.49\linewidth]{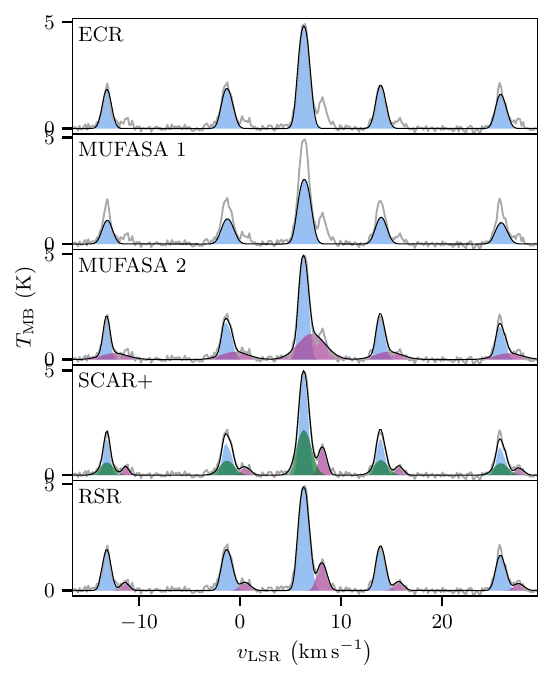}
    \caption{\label{fig:RR_Required}Two pixels which exemplify the need for the additional \ac{rsr} model. In each of the two panels, one spectrum is shown five times, with five different models as labeled.
    In the left panel, the two-component \mufasa{} model failed to produce a fit, and in the right, that fit is non-ideal.
    In both cases, the three-component \scarplus{} models are potentially overfitted. Instead, the spectra are best fit by two-component models following the \ac{rsr} step.
    } 
\end{figure*}

Several cases (a few percent of spectra) arose which made it clear that none of our existing models (the \ac{ecr}, the one- and two-component \ac{MUFASA} models, or our up-to-three-component \ac{SCAR} model) were good fits to the data.
Often in these cases, a poorly constrained $N$-component fit was rejected in favour of a $N$+1-component fit.
\reffig{fig:RR_Required} shows two pixels which exemplify some of the motivations for performing an additional fit, which we call the \acf{rsr}. 
In the two example cases, the spectral lines are visibly non-Gaussian, suggesting that two or more components are needed. 
But in both cases, a two-component fit with \mufasa{} failed and the three-component \scarplus{} model seems unnecessary (i.e., three components are not clearly visible in the spectra based on visual inspection).
Thus, the purpose of the \ac{rsr} step is to generate robust one- or two-component fits such that unnecessary two- or three-component fits are not selected.

To avoid cases where an $N$+1-component model is selected due to a poor constraint on the $N$-component model, we define a `separation potential' to look at the degree by which one component is spectrally overlapped by another.
This parameter lets us identify cases where two components are highly blended, and thus the spectrum might be better fit by a single component.

\begin{figure}[htb!]
    \centering
    \includegraphics[width=\linewidth]{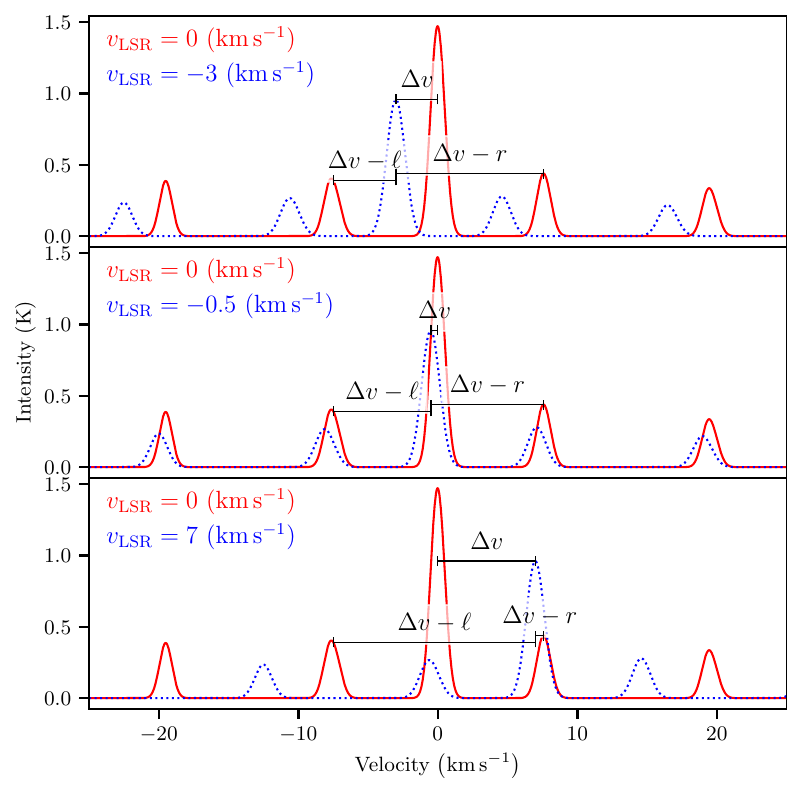}
    \caption{\label{fig:OverlapPotential_ZeroWidth}
    Examples of calculating the separation potential of two components with \refeq{eq:zero-width-overlap-potential}.
    The solid (red) component is always set at the same velocity, while the dotted (blue) component is placed differently by panel.
    \textit{Top:} two well-separated components. 
    \textit{Middle:} components which overlap in the central group.
    \textit{Bottom:} spectrum where the central group of one component overlaps the satellite group of another.  Values for the SP are given in Table \ref{tab:OverlapPotential_ZeroWidth}.
    }
\end{figure}

We define a velocity difference between components, $\Delta v = \abs{\vlsrsub{j} - \vlsrsub{i}}$ with $i$ indexing the brighter component, such that for a simplified spectrum with zero-width emission lines, we can define the separation potential in\kms{} as
\begin{equation}\label{eq:zero-width-overlap-potential}
    \text{SP} = \min{\Big[ \Delta v, \abs{\Delta v-\ell_i}, \abs{\Delta v-r_i} \Big]} ,
\end{equation}
where $\abs{\Delta v-\ell_i}$ and $\abs{\Delta v-r_i}$ are the velocity differences from the central hyperfine group of the weaker ($j^\text{th}$) component to the inner left and right hyperfine groups of the brighter ($i^\text{th}$) component.
\reffig{fig:OverlapPotential_ZeroWidth} showcases how this separation potential characterizes the velocity difference between the two components, capturing not only cases where the two central hyperfine groups overlap, but where the central group of one component overlaps an inner satellite of the other.
As can be seen in \reffig{fig:OverlapPotential_ZeroWidth}, lower values of SP indicate a higher potential for overlap between the two components.
\begin{table}[htb!]\centering\caption{\label{tab:OverlapPotential_ZeroWidth}Distances labeled in \reffig{fig:OverlapPotential_ZeroWidth}}
    \input{Tables-OverlapPotentialValues}
\end{table}

For a full model which incorporates the widths of the two components, we define 
\begin{equation}\label{eq:has-width-overlap-potential}
    \begin{split}
        \text{SP} = \min \left[\right.&\abs{\abs{\Delta v\pm \vdobssub{j}}-\vdobssub{i}}, \\
                              &\abs{\abs{\Delta v-\ell\pm \vdobssub{j}}-\vdobssub{i}},\\
                            &\abs{\abs{\Delta v-r\pm \vdobssub{j}}-\vdobssub{i}}\left. \right].
    \end{split}
\end{equation}
In the case of two velocity components, we only need to compute the separation potential between them.
If the pixel has three components, however, the separation potential is computed between the brightest component and each of the other components (i.e., between $i$ and $j$, and between $i$ and $k$).

After running the \scarplus{} pipeline, a pixel contains either one, two, or three velocity components.  The \ac{rsr} routine is performed only on pixels with velocity components that have separation potentials  \kms{$<1$}.
For example, in the left panel of \reffig{fig:RR_Required}, the central blue component of the \scarplus{} row was removed, as it overlapped greatly with both the green and purple components which are then used in the \ac{rsr}. (In the right panel, it is the central purple component of the \scarplus{} row that is removed.)
The SP threshold of \kms{1} was chosen by experimentation on a subset of clouds, examining pixels which appeared to require a reduced-component refit, and selecting a value which preferentially captured the majority of these pixels.

Each captured pixel is fitted according to the parameter limits in \reftab{tab:DefaultParameterLimits} where the parameters of the retained components are used as guesses. 
Both the blanket \ac{snr} cut of 3 and the multi-component \ac{snr} cut of 7 are applied.
Models are then generated from the \ac{rsr} routine to be used for comparison in the statistical testing.

\subsection{Model Selection}\label{sec:statistical_testing}

\begin{figure*}[htb!]
    \centering
    \includegraphics[trim=0mm 2mm 0mm 2mm,clip=true,width=0.45\linewidth]{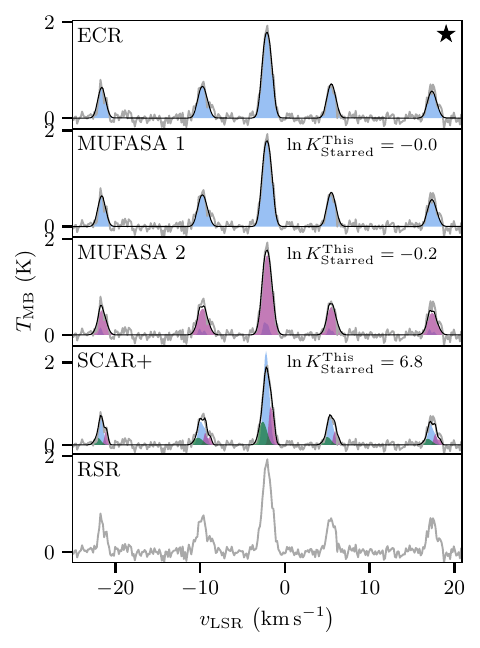}
    \includegraphics[trim=0mm 2mm 0mm 2mm,clip=true,width=0.45\linewidth]{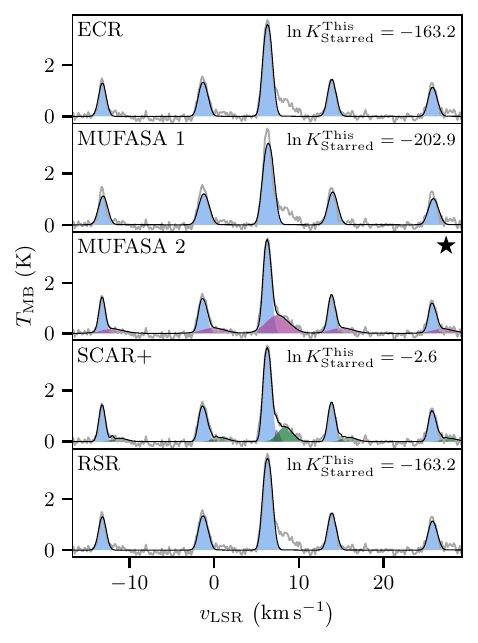}
    \includegraphics[trim=0mm 2mm 0mm 2mm,clip=true,width=0.45\linewidth]{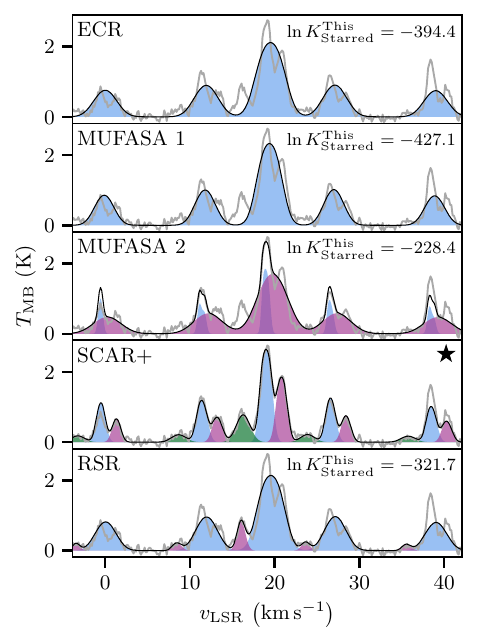}
    \includegraphics[trim=0mm 2mm 0mm 2mm,clip=true,width=0.45\linewidth]{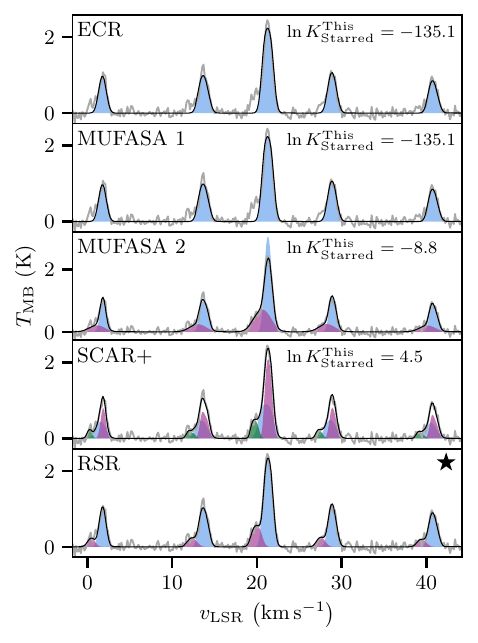}
    \caption{\label{fig:AllModelsExample}
    Four example pixels comparing all fitted models, where the best-fit model is marked with a star.
    The top-left panel does not include an \ac{rsr} model as a one-component model had already been deemed ideal.
    While in these cases the \ac{ecr} and single-component \mufasa{} models broadly agree, this is not universal (e.g., \reffig{fig:RR_Required}).
    }
    \end{figure*}

The \scarplus{} routine generates multiple spectral fits with different numbers of velocity components, ranging from one to three.
\reftab{tab:FittingRoutineComponents} summarizes the different models that can be selected, and \reffig{fig:AllModelsExample} compares each of these models as applied to several example pixels where a different model is considered the best fit.

\begin{table*}[htb!]\centering\caption{\label{tab:FittingRoutineComponents}Comparison of Models}

    \begin{tabular}{lcl} \toprule
        \th{Fitting Routine} & Components & \th{Comments} \\ \midrule
        \ac{ecr} & 1 & Modification of the original fit by \citetalias{keownKFPAExaminationsYoung2019}.\\
        \mufasa{} 1 Component & 1 & Run on all pixels in mask at once.  \\
        \mufasa{} 2 Component & 2 & Run on all pixels in mask at once. \\
        \scarplus & 1, 2, or 3 & Fit in individual subgrids. \\
        \ac{rsr} & 1 or 2 & Run only for qualifying pixels. \\
        \bottomrule
    \end{tabular}\end{table*}

As the \ac{rsr} model is a special case, we first discuss the general method for determining a best-fit model, and in the following section discuss the method for evaluating whether the \ac{rsr} model should be selected over the initial best-fit model.

\subsubsection{Primary Model Selection}\label{sec:primaryModelSelection}

We initially consider the \ac{ecr}, \mufasa, and \scarplus{} results and use statistical tests to help determine which model is superior.
We use the \acl{aicc} \citep[\acsu{aicc};][]{akaikeNewLookStatistical1974, sugiuraFurtherAnalysisData1978} following the approach in \citet{chenVelocityCoherentFilamentsNGC2020}.
The \ac{aicc} is a modification of the reduced $\chi^2$, which more heavily penalizes complex models, and thus requires that more complex models have better results in order to be accepted.
The \ac{aicc} is calculated by
\begin{equation}
    \text{AICc} = 2p + n\ln{\left(\frac{\text{RSS}}{n}\right)} + \frac{2p(p+1)}{(n-p-1)},
\end{equation}
where $p$ is the number of parameters (4 per component), $n$ is the number of data points (length of the spectral axis), and RSS is the residual sum of squares $\text{RSS} = \sum (r_i^2)$ where $r_i$ are the residuals of the model.
Lower values of \ac{aicc} indicate a more statistically preferred fit.

To select a preferred model, we used a log-likelihood value $\ln{(K)}$:
\begin{equation}
    \ln{\left(K_B^A\right)} = \frac{\text{AICc}_B - \text{AICc}_A}{2}.
\end{equation}
If the value of $\ln{\left(K_B^A\right)}$ is positive, Model A is preferred over Model B (the converse is true for a negative log-likelihood), with the strength of that preference increasing as the value moves farther from zero.
In general, $\ln{\left(K_B^A\right)}>5$ is considered evidence that Model A is favoured over Model B and a value of $\ln{\left(K_B^A\right)}>20$ is strong evidence \citep{burnhamMultimodelInferenceUnderstanding2004}.
In all cases, we define Model A as the more complex model.

Models were selected as they were generated, and so each new model is compared only to the current best-fit model, rather than to every model generated up to that point.
When comparing two models, there are several cases with different levels of model complexity.  As such, we adopted log-likelihood thresholds following each case:
\begin{itemize}
    \item For models of equivalent complexity (e.g., the \ac{ecr} generated from \citetalias{keownKFPAExaminationsYoung2019} and the single-component \mufasa{} model),  we use a threshold of $\ln(K)=0$ to determine a preferred model.
    \item For an $N$-component model compared to an $N$+1-component model, we required that $\ln{\left(K_N^{N+1}\right)}>5$ in order for the $N$+1 components to be selected. This value was also used by \citet{chenVelocityCoherentFilamentsNGC2020}, and ensures that more complex models are only selected when they are well preferred.
    \item For an $N$-component model compared to an $N$+2-component model, we increased the threshold for accepting the more complex model to $\ln{\left(K_N^{N+2}\right)}>10$.
    \item For any spectral model to be selected over a noisy model (no emission), the model must be strongly preferred, i.e., it must have $\ln{\left(K_\text{noise}^A\right)}>20$.
\end{itemize}

\subsubsection{Selecting Reduced \acs{SCAR} Refit Models}\label{sec:selecting-rsr}

The \ac{rsr} models are not compared against models of equivalent complexity but instead against the best-fit model from the primary selection outline in the previous subsection. 
By definition, the primary best-fit model has at least one additional component compared to the \ac{rsr}.

The flow chart in \reffig{fig:FlowchartDecisionTree} explains the decision tree we use for model selection following the \acl{rsr}.
We first evaluate the log-likelihood value of the \ac{rsr} model ($\text{AICc}_A$) as compared to the primary best-fit model ($\text{AICc}_B$).
If this indicates a preference to the \ac{rsr} (i.e., $\ln(K)\geq 0$), we select the \ac{rsr} model.
As the RSR model is defined to have fewer components than the previous best-fit model, this is in line with our usual method as described in \refchap{sec:primaryModelSelection}.

\begin{figure}[htb!]
    \centering
    \input{Figures-FlowchartDecisionTree}
    
  \caption{\label{fig:FlowchartDecisionTree} Decision tree showing our choice hierarchy for determining the acceptance or rejection of the \ac{rsr} model over the primary best-fit model.
    This decision tree was created step-by-step and is applied pixel by pixel. 
    }
\end{figure}

In the case that the \ac{rsr} model is not statistically preferred, we then  evaluate the zero-width separation potential between the removed component using \refeq{eq:zero-width-overlap-potential}.
If the separation between velocities in the two components is \textgreater~2 channel widths (\kms{\textgreater~0.28}), we consider the components to be well enough separated such that the primary best-fit should be favoured over the \ac{rsr} fit. 
We therefore return to the log-likelihood, and in cases where $\ln(K)<-50$, i.e., we have a very strong preference to the primary model, we accept the primary best-fit model over the \ac{rsr}.   

\begin{figure}[htb!]
    \centering
    \includegraphics[width=\linewidth]{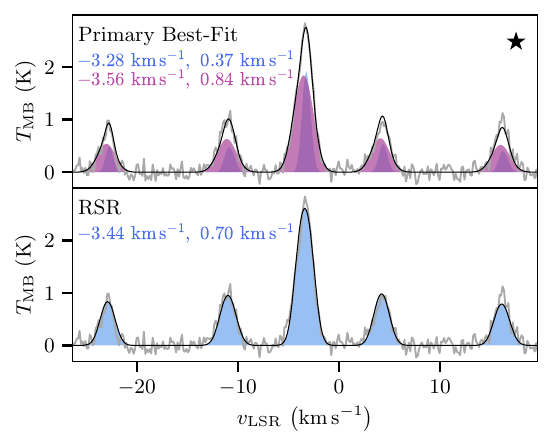}
    \includegraphics[width=\linewidth]{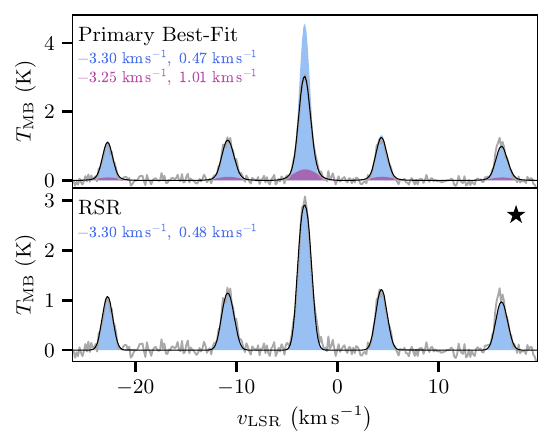}
    \caption{\label{fig:KinematicChangeExamples} Examples of pixels with large  (top) and small (bottom) values of $\Delta k$. 
    The relevant model parameters are included on the plot, in units of \mbox{km s$^{-1}$}.  The starred panels indicates the final selected model. \textit{Top:} the purple component is kept and re-fit in the \ac{rsr}, resulting in $\Delta k=4.64$. \textit{Bottom:} the blue component is kept for the \ac{rsr}, with $\Delta k=0.07$.
    }
\end{figure}

If all of the above tests fail to identify a preference for either the \ac{rsr} or the primary best-fit model, we apply a final test to determine whether or not the velocity parameters change significantly during the \ac{rsr}. 
The motivation for this final test is that a large change in the velocity parameters indicates that the extra component has an impact on the fitted spectra and therefore cannot be ignored.
To evaluate the change in kinematic information, we define a dimensionless value $\Delta k$, 
\begin{equation}\label{eq:kinematic_change}
\begin{split}
    \Delta k = &\frac{1}{sN}\sum_i^N\abs{\vlsrtext{RSR,$i$}-\vlsrtext{primary,$i$}}\\
    &+\frac{1}{2sN}\sum_i^N\abs{\vdobssub{\text{RSR},i}-\vdobssub{\text{primary},i}},
\end{split}
\end{equation}
where \kms{$s=0.14$} is the spectral channel width and $N$ is the number of components included in the \ac{rsr} for that pixel. 
The factor of 2 in the centroid velocity term adds additional weight to that term, as we consider a change in centroid velocity to be particularly important.
The parameter $\Delta k$ quantifies how much the centroid velocity and velocity dispersion change between the \ac{rsr} and primary best-fit models averaged over all components which were used in the \ac{rsr}.  For example, if the \ac{rsr} parameters were identical to the primary model parameters, the resulting $\Delta k$ would be 0.
We do not include any removed components  in the $\Delta k$ calculation as we are essentially evaluating whether those removed components affected the kinematics.
\reffig{fig:KinematicChangeExamples} shows examples of both a large and small $\Delta k$.
If the average (over all components) change in these two quantities falls within $\Delta k < 2.5$, we say that the change in the kinematic information is insignificant, and accept the \ac{rsr}.
If the change in kinematic information is $\Delta k \geq 2.5$, then we keep the primary best-fit model.

Following the selection of the final best-fit model, we examined a variety of spectra, including those with the brightest residuals, and concluded that a fourth velocity component was not obviously present.
We discuss a validation of the entire \scarplus\mufasa{} pipeline, including the \ac{rsr} and model selection in \refapp{app:validation}.

\section{Results}\label{sec:results}
\subsection{Data Products}
With this data release, we provide the pertinent final data products from the multi-component fit pipeline as well as the \ac{SCAR} code.
Intermediary and additional products (i.e., model cubes, products from one- and two-component fitting, etc.) can be obtained by running the \ac{SCAR} code.

Parameter cubes (\formatSoftware{parcubes}) contain the fitted parameter values. 
The \formatSoftware{parcubes} are named based on a CLOUD identifier, where CLOUD indicates both the larger GMC  and associated subregion as identified in Tables \ref{tab:GMC-Summary} and \ref{tab:SubcloudCoordinates}.
Each \formatSoftware{parcube} contains the fitted parameter values in different frames.  The fitted parameters are centroid velocity (\verb|vLSR_i|), velocity dispersion (\verb|vDISP_i|), excitation temperature (\verb|TEX_i|), and optical depth (\verb|TAU_i|) for each component (\verb|i=1,2,3|); as well as the associated errors (\verb|e_vDISP_i|, etc.).
The uncertainties presented in our parameter cubes are those reported by \pyspeckit{} as products of the fitting process.  
All data products use NaN (not a number) values as markers for locations where either data were not computed (as in parameters at a pixel which was masked out), a fit was attempted but not found, or the few (\textless 10) cases where a fit appears to incorrectly fit the data.  
This information is also given in a README file included on the \ac{canfar} system, and further information about the data products can be found in the FITS headers of the provided files.

We stress that there is no inherent meaning to the identification of component 1/2/3. 
The component order represents their order in terms of the radiative transfer and only impacts the results in the cases of optically thick emission.
As discussed in Section \ref{sec:ScarPlusConsiderations} and  exemplified in \reffig{fig:ComponentSwitchingExample}, the component order can switch between adjacent pixels and has no continuity across the maps.  As such, care will be required when identifying coherent structures from these parameter cubes.
A forthcoming paper \inprep{Withers} will apply structure finding algorithms to identify velocity-coherent structures and link the velocity components.

\subsection{Multicomponent Fitting Statistics\label{sec:fittingStats}}

\reftab{tab:ComponentStatisticsTabular} lists the percentage of pixels with each number of velocity components.
W3 West contained no multicomponent pixels, though it also had by far the fewest fitted pixels.
By contrast M17, NGC 2264, and W48 had the highest fractions of multi-component pixels, with $>15\percent$ of fitted pixels needing two or more components.
We note that due to our rejection of any multicomponent model where all components had \ac{snr}\textless~7, our statistics may be biased towards a slightly more conservative number of components.

\begin{table}[htb!]
    \centering\caption{\label{tab:ComponentStatisticsTabular}Prevalence of multicomponent pixels by cloud}
    \input{Tables-ComponentStatistics}    
\end{table}

\begin{figure}[htb!]
    \centering
    \includegraphics[width=\linewidth]{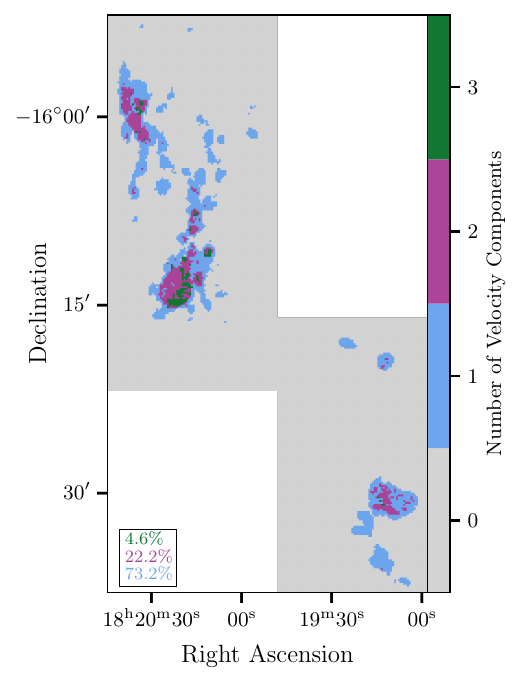}
    \caption{\label{fig:ncmap} Example of the final number of velocity components selected by our fitting routine for M17.
    }
\end{figure}

While the relative number of multiple component pixels is relatively small, these pixels are generally clustered together.
For example, \reffig{fig:ncmap} shows the locations where one, two, and three components were fitted in M17 (similar component maps for the other \keystone{} clouds are provided in \refapp{app:ncmaps}). The regions where multiple (2 or 3) components are necessary are in the dense regions where the \nhhh{} Moment 0 values are highest, with values $\gtrsim 10$ K km s$^{-1}$. Similar thresholds for multiple components are seen for the other clouds.  
Conversely, W3 West, which has no cases of multi-component fits, has the lowest Moment 0 emission ($< 7$ K km s$^{-1}$). 
Assuming to first order that the Moment 0 emission correlates with column density, N(\nhhh{}), and hence gas density, then the detection of multiple velocity components in high Moment 0 regions would imply that the multiple components are found primarily in higher density environments.
Thus, the multi-component fitting is critical to accurately measure the gas dynamics in the regions where star formation is most active in these \acp{gmc}.  We discuss this correlation between multiple components and high-mass star formation more in Section \ref{dis:HMSF}.

\begin{figure}[htb!]
    \centering
    \includegraphics[width=\linewidth]{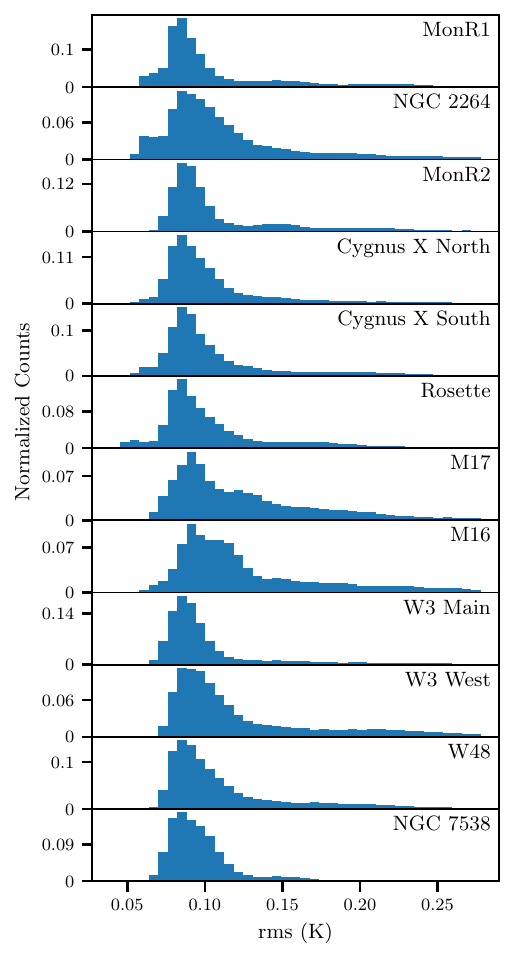}
    \caption{\label{fig:rms_histograms}
    The rms values resulting from our multicomponent fit to the \nhhh{} data, clouds are ordered by increasing distance.
    }
\end{figure}

We calculated the rms for each spectrum as described in \refsec{sec:TwoComponentFitting}, using the channels where our spectral models had intensities $\leq$0.0125~K.  \reftab{tab:GMC-Summary} lists the median noise value for each cloud. These values are slightly lower than reported by \citetalias{keownKFPAExaminationsYoung2019}, and this difference is due to our use of spectrally smoothed data cubes and our fitting of additional velocity components.  \reffig{fig:rms_histograms} plots the range of rms values for each cloud.

\reftab{tab:medianSNR} shows that the \ac{snr}, measured as the ratio of the peak brightness of our model to the rms value, increases with the number of components selected for a given pixel.
Thus we see that multiple components are found at sites with brighter emission, where we expect dense gas to be located, which is a precondition for star formation.
Alternatively, multiple components may be biased to appear in brighter regions because of our requirement that multiple-component models contain at least one component with \ac{snr} $\geq 7$. We discuss a correlation between multiple components and S/N more in Section \ref{dis:HMSF}.
\begin{table}[htb!]\centering
	\caption{\label{tab:medianSNR}Median \acs{snr}
    }
    \input{Tables-ComponentSNRs}
\end{table}

Finally, \reftab{tab:ModelStatisticsTabular} breaks down the frequencies with which each of our models is selected. 
All the models have the capacity to fit one velocity component, and several models can fit two components.  For simplicity, we combine all the \keystone{} regions and examine the frequency of each model as a function of fitted pixels.  For the single-component pixels, the ECR model generally produces the best fit, although a non-negligible fraction of single-component pixels were better fit using the other models. For the two-component pixels, \mufasa{} performs best, but some \scarplus{} and RSR models yield better results.  The variety of model selection suggests that no single model can fully reproduce the \nhhh{} spectra in the \keystone{} clouds, particularly in the regions with the most \nhhh{} emission and most active star formation.

\begin{table}[htb!]\centering\caption{\label{tab:ModelStatisticsTabular}Frequency of Model Selection
    }
    \input{Tables-ModelStatistics}
\end{table}

\subsection{Notable Cases During Fitting}

\begin{figure}[htb!]
    \centering
    \includegraphics[width=\linewidth]{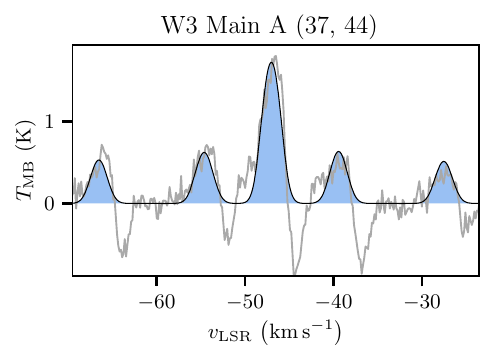}
    \caption{A spectrum showing an ammonia absorption feature and our best-fit model, where we enforce a single-component fit.
    This absorption feature is possibly a result of the frequency switching during observations capturing a unintended \nhhh{} emission.
    }
    \label{fig:W3_Absorption}
\end{figure}

One region in W3 Main around (02:27:04.5, 61:52:30) in subcloud A exhibited absorption features matching the \nhhh{} spectrum in a small subset of pixels. \reffig{fig:W3_Absorption} shows an example of a spectrum in this region.
This absorption feature nearly directly overlapped the \nhhh{} emission spectrum, which interfered with our ability to robustly fit the data.
As a result of this absorption, we masked these pixels in early stages of the pipeline, and where this was not the case, a maximum of one component is manually enforced.

\begin{figure}[htb!]
    \centering
    \includegraphics[trim={0 0 15cm 0},clip,angle=270, origin=c,width=\linewidth]{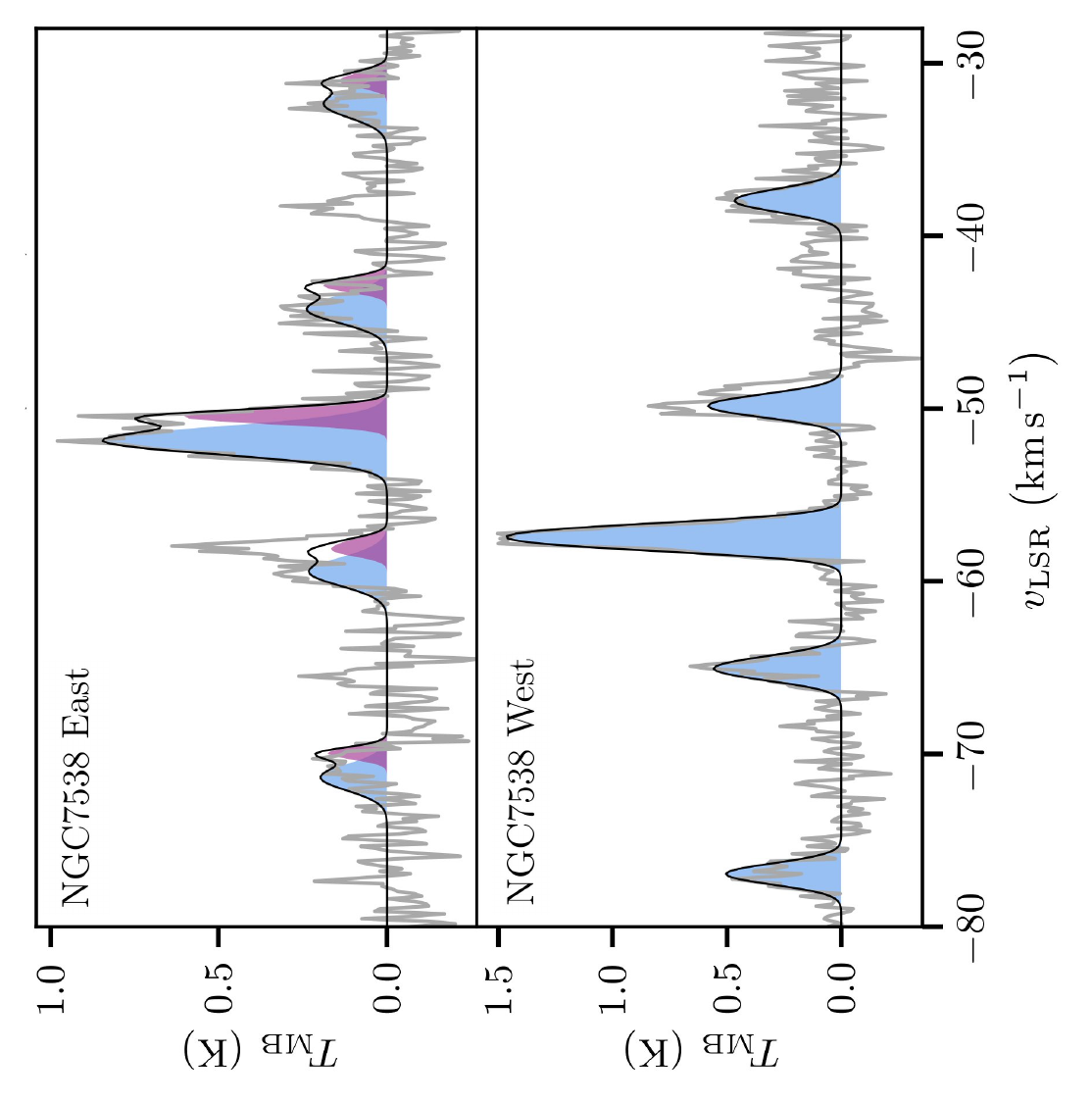}
    \caption{Two fitted spectra from NGC 7538 East (top) and West (bottom), sharing the same velocity axis  showing how the central \nhhh{} hyperfine group from one cloud overlaps with a satellite \nhhh{} group from the other.  Both panels show unfitted emission due to the other cloud.  For the upper panel, the Eastern cloud shows excess emission with the blue-inner satellite group at \kms{$\approx -57$}, which is characteristic of the centroid velocity of the Western cloud. Similarly for the lower panel, the Western cloud shows excess with the red-inner satellite group at \kms{$\approx -50$}, which corresponds to the velocity of the Eastern cloud.
    }
    \label{fig:NGC7538_Overlap}
\end{figure}

NGC 7538 has Eastern and Western subclouds with line-of-sight velocities of \kms{$\approx~52$} and \kms{$\approx~57$} respectively \citep[e.g.,][]{rigby2025}.
This centroid velocity difference is roughly equal to the velocity separation between the central hyperfine group and the inner satellites (e.g., see lower panel of \reffig{fig:OverlapPotential_ZeroWidth}). As a result, NGC 7538 has pixels where the central hyperfine group from one cloud overlaps with an inner satellite group from the other cloud. 
\reffig{fig:NGC7538_Overlap} shows two example pixels highlighting this situation where an inner satellite group of the `primary' cloud is overlapped by the central hyperfine group of the `secondary' cloud.
Fitting this emission with an additional \nhhh{} component proved non-trivial as the emission is dim enough that the satellite groups of the secondary cloud are not consistently identifiable.\footnote{As the main satellite of these unfitted components often overlaps with an inner satellite of the brightest fitted component, our parameter limits for the three-component fit (Step 8 of \scarplus{}) often excludes this third component, forcing these pixels to be fit with a maximum of two velocity components. See \refapp{app:trifitparlims} for details of how these parameter limits are determined.}
We note that our final data products do not appear to have notable fitting errors as a result of this overlap.

\section{Discussion}\label{sec:discussion}

\subsection{Multiple Components and High-mass Star Formation}\label{dis:HMSF}

Figure \ref{fig:ncmap} shows that the pixels with multiple velocity components tend to be clustered together (see also Appendix \ref{app:ncmaps}).  To examine the physical meaning of these multiple velocity components, we correlate them with tracers of gas (column) density.  We use our \nhhh{} Moment 0 maps and the N(H$_2$) column density map produced from \emph{Herschel} data from the \ac{hobys} survey \citep{motteInitialHighlightsHOBYS2010,motteHighMassStarMassive2018} as proxies for density.  For the N(H$_2$) data, we first re-grid the map to match the pixel scale of the \nhhh{} data.  Note that the N(H$_2$) maps at 36\arcsec\ resolution are comparable in resolution to our \nhhh{} data.  Nevertheless, the \emph{Herschel} column densities are derived from dust and as such, they do not differentiate between multiple components like the Moment 0 maps.

To evaluate how the multiple components correlate with  density, we break up the clouds into 10 equally-sized (roughly 5000 pixels) percentile bins in either Moment 0 or N(H$_2$) to equally identify low to high densities across all the KEYSTONE clouds irrespective of GMC.
Tables \ref{tab:ComponentStatisticsMom0} and \ref{tab:ComponentStatisticsNH2} give the prevalence of 1, 2, or 3 velocity components and the median velocity dispersion in the 10 percentile bins for the \nhhh{} Moment 0 and N(H$_2$) maps, respectively. 
Table \ref{tab:ComponentStatisticsMom0} also shows the median peak S/N for the pixels in the percentile bins with 1, 2, and 3 components. 
In both cases, we find that multiple components are typically associated with higher density material. 
Over 90\% of pixels in the lower ($<$ 60th percentile) bins are best fit with one velocity component, whereas in the top ($>$ 90th percentile) bins, over 50\% of pixels are best-fit with 2 or 3 velocity components.  

\begin{table*}[htb!]
\centering
\caption{\label{tab:ComponentStatisticsMom0} Gas Properties in \momzero{} Bins
}
\input{Tables-ComponentPercentilesMom0}

\tablecomments{For the S/N and velocity dispersion, the quantities given are the median values and the 1 $\sigma$ standard deviation (evaluated using the 16th and 84th percentile levels from the distribution) for the column density percentile range.  }
\tablenotetext{a}{We include the single-component (ECR) results and the \scarplus{} results separately to highlight the significance from multiple component fitting. Note, that the number of values used to calculate the median velocity dispersion between the \ac{ecr} and \ac{SCAR} models differ slightly since the fitting criteria are not identical and the \scarplus{} fits have multiple components. The highest difference in the number of pixels used to calculate the median is \percent{$< 1$}.}
\end{table*}

\begin{table*}[htb!]\centering\caption{\label{tab:ComponentStatisticsNH2}Gas properties in N(\hh) Bins
    }
\input{Tables-ComponentPercentilesNH2}\tablenotetext{a}{See notes on velocity dispersion in \reftab{tab:ComponentStatisticsMom0}.}\end{table*}

Broadly, these results show that multiple velocity components are typically found toward higher density regions as traced by \nhhh{} Moment 0 ($\gtrsim 5$ K\kms{}) and dust-derived column density ($\NHH\ \gtrsim 10^{23}$ \cden), which suggests that the gas kinematics may be changing uniquely at these density thresholds. 
Prior \emph{Herschel} studies found that column densities $\gtrsim 10^{23}$ \cden\ in GMCs are typically associated with hubs, ridges, and dense clumps, which are typically associated with high-mass star formation as evidenced from embedded massive young stellar objects (MYSOs), young forming clusters, and ultra-compact HII regions \citep[e.g.,][]{Hill2011, Schneider2015, motteHighMassStarMassive2018,Schisano2020,pinedaBubblesFilamentsCores2023}.  Moreover,  \citet{schneiderUnderstandingStarFormation2022} constructed probability density functions (PDFs) using column density (extinction) maps derived from \emph{Herschel} data in the HOBYS survey and found that the massive star forming regions often required two power-law tails compared to a single power-law tail for the lower-mass regions. The second power-law tails were typically associated with material with $A_V \gtrsim 55$ mag ($\gtrsim 3.9 \times 10^{22}$ \cden\ or $\log\NHH \gtrsim 22.6$), which also corresponds to our highest percentile bin. The presence of a second power-law tail unique to high-mass star-forming regions suggests that there is a change in physical conditions or processes at those higher densities. 

Recent observations from the Atacama Large Millimeter/submillimeter Array (ALMA) have observed hubs, ridges, and dense clumps in high-mass star-forming GMCs on small ($\lesssim 1$ pc) scales \citep[e.g.,][]{Cunningham2023,Das2024,SandovalGarrido2025, Koley2025}.  Indeed, these observations show networks of spatially resolved subfilaments and dense cores with complex velocity features and prominent velocity gradients that would be unresolved in lower-resolution single-dish data.  Thus, the high fraction of multiple components at the highest gas column densities in the KEYSTONE data could arise from similar unresolved gas flows toward the sites of massive star formation. 

Alternatively, we may find multiple velocity components primarily in the high column density regions because those regions are also brighter in \nhhh{} emission allowing these fits to have at least one component which meets our S/N $> 7$ threshold. 
To test whether or not we may be biased to finding additional velocity components toward bright regions due to better S/N at those locations, we compare the S/N values for each component in each Moment 0 percentile bin in  Table \ref{tab:ComponentStatisticsMom0}. 
Below the 60th percentile bin, the single-component pixels tend to have S/N values that drop well below the threshold of 7 used to identify multiple components. 
Therefore, in these lower intensity regions, we are likely biased to missing additional components due to our conservative threshold needed for a robust multi-component fit.
evertheless, above the 60th percentile bin, the single-component pixels tend to have S/N ratios in excess of 7 and generally have good agreement with the multiple-component pixels.
Thus, the increasing fraction of multiple velocities at the highest densities are likely intrinsic and represent different cloud components on smaller scales within the GMCs.

In addition to finding high fractions of multiple velocity components at high gas column densities, we also find non-negligible cases of multiple velocity components in the lower density regions (e.g., $\NHH \sim 2 \times 10^{22}$ \cden).  At these column densities, we are unlikely to be tracing sites of massive star formation like hubs and dense cores, and may instead be tracing subfilaments within the larger cloud complexes.  Specifically, observations have shown that dense filaments in low-mass and high-mass star-forming regions often show networks of braided subfilaments with different velocity components that have been termed `fibers' \citep[e.g.,][]{Hacar2013,Shimajiri2019,chenFilamentAccretionFragmentation2024}.  There are two main theoretical pictures for the formation of such fibers.  The first is a gravito-turbulent model where a filament is produced via shocks from turbulent compressive flows that ultimately fray into individual fibers \citep[e.g.,][]{MacLowKlessen2004, CYChen2020}.  The second is a global hierarchical collapse model where infall motions from the larger collapsing cloud produce individual fibers that gather together into a filament under gravity \citep[e.g.,][]{RSmith2016, VazquezSemadeni2019}.   Ultimately, higher resolution data will be necessary to verify whether or not the different velocity components are fibers and if so, differentiate between these scenarios.

\subsection{Velocity Dispersion}

\begin{figure*}[htb!]
    \centering
    \includegraphics[width=\linewidth]{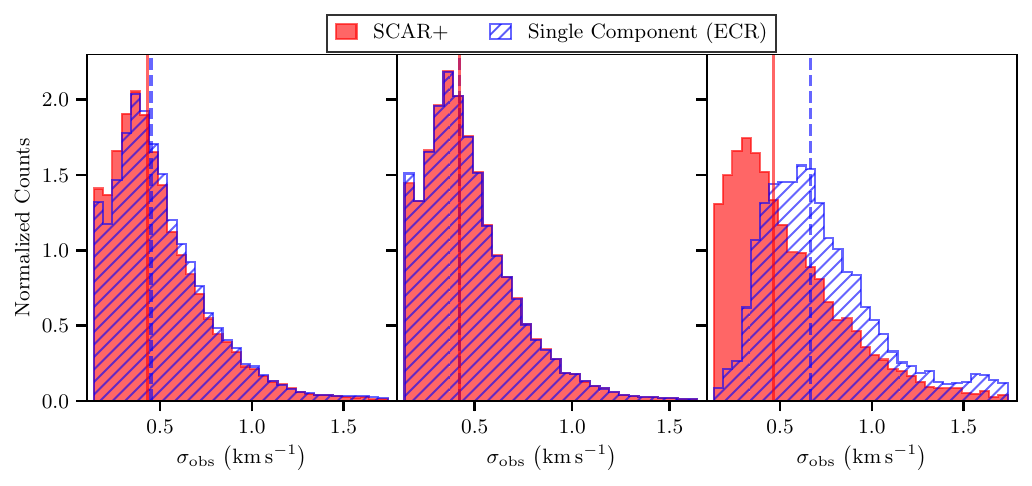}
    \caption{\label{fig:AllCloudsDispersionChange}
    Comparison of velocity dispersions between single- and multiple-component fits. Vertical lines show median values for each distribution.
    \textit{Left:} all pixels fitted in our analysis.
    \textit{Centre:} only pixels we select as single-component.
    \textit{Right:} only pixels identified as having multiple components; note how the multi-component analysis finds lower velocity dispersions than a single-component fit.
    Histograms are normalized such that the integral of each dataset is unity.
    }
\end{figure*}

\reffig{fig:AllCloudsDispersionChange} compares the distributions of velocity dispersions resulting from our multiple component fits as compared to the single-component fit from \kt{} (via the \ac{ecr}; see \refsec{sec:TwoComponentFitting}). 
The vertical lines signify the median values for each distribution. 
The median fitted velocity dispersion across all clouds from the single-component \ac{ecr} fits is \kms{0.45}, which is in good agreement with \citetalias{keownKFPAExaminationsYoung2019}.
For our multi-component fitting, the median velocity dispersion lowers slightly to \kms{0.43} across all fitted pixels.
These values are relatively similar, as expected based on the majority of pixels being fit with single-component models (see also, left panel and middle panel of \reffig{fig:AllCloudsDispersionChange}). 
Nevertheless, if we consider only the pixels for which we identify multiple components (right panel of \reffig{fig:AllCloudsDispersionChange}), the median velocity dispersion using the \scarplus{} routine is similar at \kms{0.47}, whereas the single-component fits to the same pixels from the \kt-based \ac{ecr} models yield higher values by a factor of $\sim$ 1.4 or \kms{0.67}.
The greater difference for multiple-component pixels is expected as fitting one component does not accurately capture the gas kinematics (e.g. \reffig{fig:RequiresThirdComp}).  

Tables \ref{tab:ComponentStatisticsMom0} and \ref{tab:ComponentStatisticsNH2} show the median velocity dispersions for each percentile bin of \nhhh{} Moment 0 and N(\hh), respectively, for all components.  
Broadly the ECR and \scarplus{} velocity dispersions agree well for the lower percentile bins, but deviate by $\gtrsim 10$\%\ above the 70th percentile bins.  
Nevertheless, both the single and multiple component fits show  that the median velocity dispersion increases from the lowest  Moment 0 and N(\hh) bins to the highest bins. In both cases, the median velocity dispersion increases by roughly a factor of 2 between the lowest and highest percentile bins.
As massive star formation is typically associated with the highest percentile bins (see Section \ref{dis:HMSF}), the larger velocity dispersion seen in these pixels could reflect feedback from high mass star formation or artificial broadening (e.g., due to unresolved velocity flows within the GBT beam). 
The velocity dispersion distributions in Figure \reffig{fig:AllCloudsDispersionChange} show a long tail that may be indicative of such turbulent-driven feedback or larger-scale velocity gradients, such as from collapse.

Many studies use \nhhh{} line widths to infer turbulent support in dense cores using the virial equation \citep[e.g.,][]{pattleJCMTGouldBelt2015, singhVirialRatioDirect2019, keownKFPAExaminationsYoung2019}. 
Since the multi-component fits are primarily found in the dense environments (see Section \ref{dis:HMSF}), they may be also associated with individual dense cores. 
As such, our multi-component fits have the potential to lower the measured level of turbulence and turbulent support in dense cores (e.g., by separating the motions of large-scale cloud flows from those of the dense cores), which can in turn change the stability of the cores using metrics such as a virial analysis \citep[e.g.,][]{choudhuryUbiquitousNH3Supersonic2020}.  
An examination of dense cores in these \acp{gmc} and their revised virial parameters will be presented in a follow-up study \inprep{Withers}.
  
\begin{figure*}[htb!]
\centering
    \includegraphics[width=\linewidth]{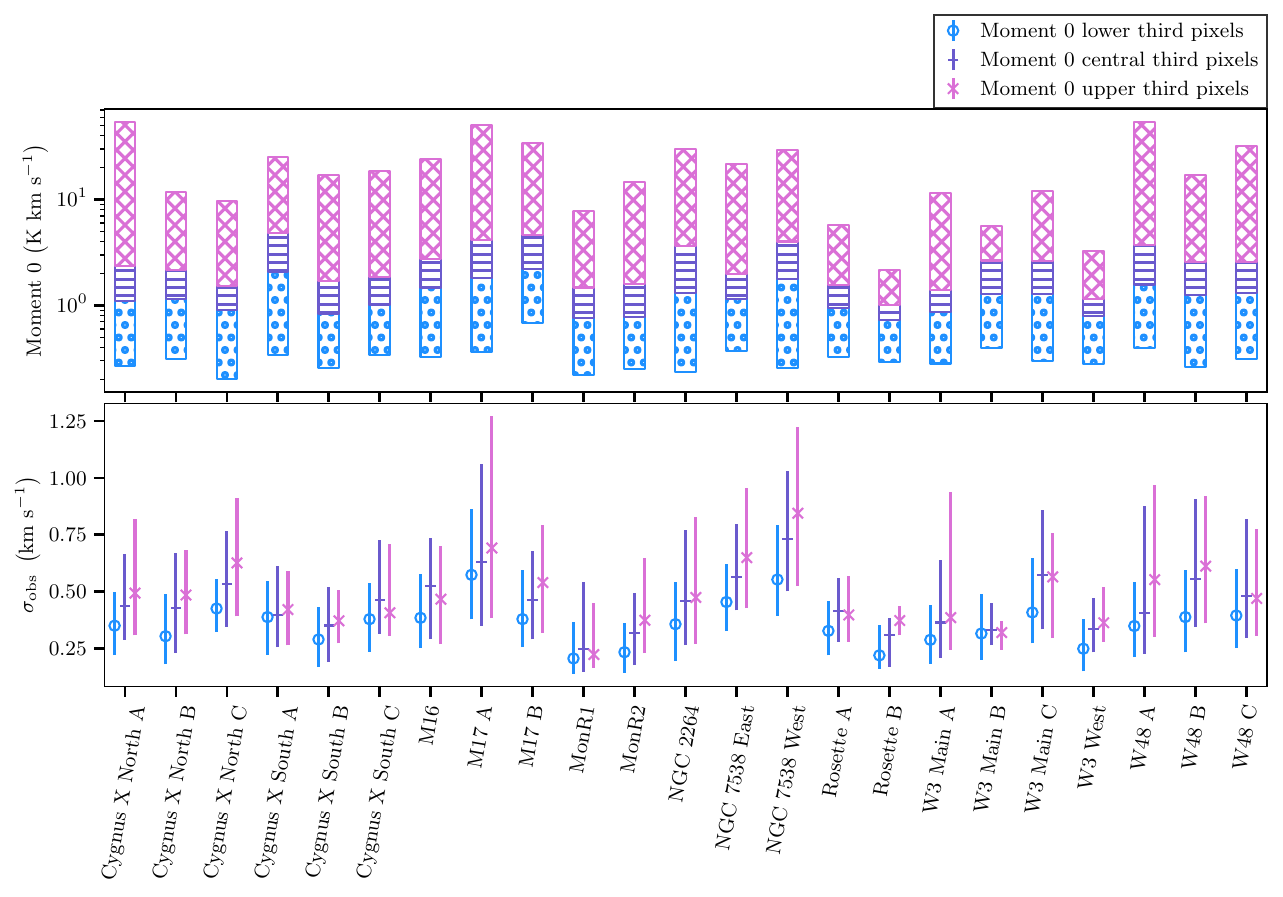}
\caption{\label{fig:VelocityDispersionByCloud}
    Comparison of \nhhh{} \momzero{} and velocity dispersion within each subcloud. 
    \emph{Upper panel:} Bar graphs showing the range of \momzero{} values per subcloud with different colours denoting the ranges for pixels in the lower third, middle third, and upper third of \momzero{}. \emph{Lower panel:} Points give the median velocity dispersions and 1 $\sigma$ ranges for the pixels which fall into the above \momzero{} bins. The 1 $\sigma$ error bars are measured from the 16th and 84th percentiles.
    }
\end{figure*}

While the median velocity dispersion increases with density, even in the high density bins there are some pixels with relatively narrow line widths.  Tables \ref{tab:ComponentStatisticsMom0} and \ref{tab:ComponentStatisticsNH2} show that both the median velocity dispersions from the \ac{ecr} and \scarplus{} fits have large $1 \sigma$ values (which are measured from the 16th and 84th percentiles).  
These large ranges are not unsurprising as the clouds themselves are different environments.  Figure \ref{fig:VelocityDispersionByCloud} compares the median velocity dispersion per cloud or subcloud.  To ensure we have enough pixels to measure the median, we break up each cloud and subcloud into bins representing the lower third, middle third, and upper third of \nhhh{} Moment 0.  The upper panel shows the distribution of Moment 0 in each bin and the lower panel shows the corresponding median velocity dispersion.  The clouds are presented alphabetically.  Indeed, Figure \ref{fig:VelocityDispersionByCloud} shows that that some clouds like M17 A have very broad lines whereas other clouds like NGC 2264 have relatively narrow lines even though both clouds have similar Moment 0 values (see Figure \ref{fig:VelocityDispersionByCloud}).  Some clouds like Cygnus X A and NGC 2264 show a wide range of Moment 0 values, whereas other clouds like Rosette B and W3 West have lower and narrower ranges of Moment 0.

In addition to large variation between the clouds, Figure \ref{fig:VelocityDispersionByCloud} shows variations between subclouds. 
For example, M17 A has higher velocity dispersions compare to M17 B by $\sim 50$\% and  W3 Main C has substantially higher (almost a factor of 2) velocity dispersions compared to W3 Main B when comparing the highest Moment 0 pixels.  
In some cases, this may be a matter of statistics: the `C' subcloud of W3 Main contained 1,896 pixels, while the `B' subcloud had only 257 fitted pixels.
Other instances, such as the wide linewidths on M17 A compared to M17 B, could potentially indicate environmental differences in the subclouds.  M17 A is impacted by an expanding HII bubble and shows evidence of MYSOs, whereas M17 B is more shielded and quiescent \citep[e.g.,][]{Nguyen-Luong2020,Lim2020}.
Since the subclouds were chosen based on where cloud structure is located according to \momzero{} maps,  further evaluation is needed to correlate any potential differences that might exist within these \acp{gmc} with cloud environment.

\subsection{Anomalous Hyperfine Emission}\label{sec:ahe}

\begin{figure}[htb!]
\centering
    \includegraphics[width=\linewidth]{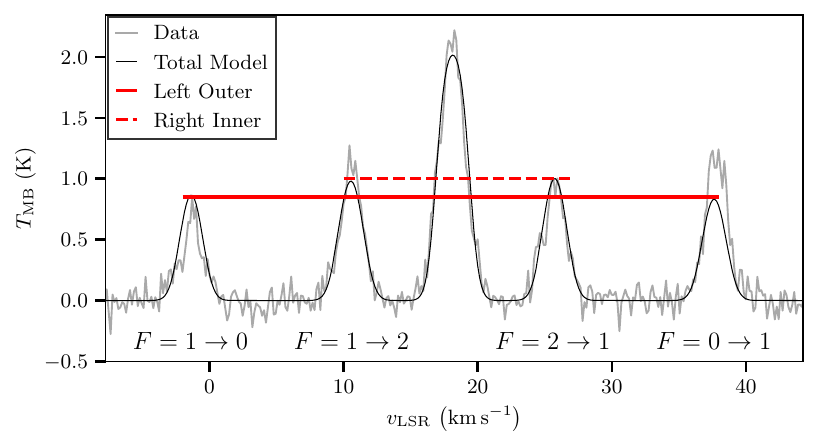}
    \caption{\label{fig:AnomalousSatellites}
    Spectrum from M17 showing the presence of \ac{ahe}. The grey spectrum shows the observed data and the best-fit fitted model, assuming \ac{lte}, is shown in black. 
    Annotations note the major hyperfine transition groups, which are further split by magnetic interactions to form the 18 hyperfine components. 
    Horizontal red lines highlight the difference in heights within the inner and outer satellite groups. In \ac{lte}, the hyperfine groups should have the same heights, but in this spectrum, the $F=0\rightarrow1$ group (outer red group) is noticeably taller than the $F=1\rightarrow0$ group (outer blue group).
    }
\end{figure}

While emission of the \nhhh{} transition has advantages in being well identifiable and easily characterized, the line ratios can deviate from standard expectations from \ac{lte}.
Initial detection of \ac{ahe}\footnote{Also called the \ac{hia}.} was reported by \citet{matsakisAnomalousAmmoniaAbsorption1977}, where the \nhhh{} transition seen in absorption had satellite group line heights which differed from the expected \ac{lte} values by a few percent.
Follow-up studies \citep[e.g.,][]{stutzkiNH3HyperfineIntensity1984, camarataNH3HyperfineIntensity2015} found that unequal line heights are prevalent throughout \acp{gmc} in star-forming regions, but fail to identify any clear link between the presence, type, or degree of anomaly with any other physical characteristic.
\reffig{fig:AnomalousSatellites} shows a particularly clear example of \ac{ahe} consistent with predictions by \citet{matsakisAnomalousAmmoniaAbsorption1977}.

We see throughout the \keystone{} clouds cases of clear \acl{ahe}, primarily in the outer satellites (e.g., Figures \ref{fig:RequiresThirdComp} and \ref{fig:AnomalousSatellites}), where the anamolous emission is typically $F=0\rightarrow1 > F=1\rightarrow0$ (see \reffig{fig:AnomalousSatellites}). 
Quantifying the occurrence and intensity of \ac{ahe} in the \keystone{} clouds is beyond the scope of this paper, particularly as the multicomponent nature of the emission requires careful fitting with excellent priors in order to differentiate anomalous emission from blended emission.
As \ac{ahe} affects only relative line heights and not their widths or center velocities, its presence is not expected to result in substantial errors in our fitting algorithms with regard to the kinematic parameters. 
In cases where \ac{ahe} is significant, our fitting results for $\tex$ and $\tau$ will be affected, however.
Nevertheless, based on initial probing of the results, \ac{ahe} generally impacts only small, dense regions in the \ac{KEYSTONE} \acp{gmc}, with the majority of the cloud area being well-fitted.
A more complete analysis of both the extent and degree of \ac{ahe} in these regions are forthcoming (Plant et al. in preparation).

\section{Summary}\label{sec:summary}
In this paper, we build on the work of \citet{chenVelocityCoherentFilamentsNGC2020} by developing \scarplus, an iterative method to fit the \nhhh{} \oneone{} spectrum with up to three velocity components.
As an extension to \mufasa, \scarplus{} incorporates radiative transfer and selects the best-fit number of components through a combination of statistical tests and physicality checks. 
\scarplus{} also has the potential to be extended beyond three components. 
The code has been made public, with the most recent version available on \textit{Zenodo}\footnote{\url{https://zenodo.org/records/23487470}}. 

We applied \scarplus{} to the \keystone{} dataset of 11 \acp{gmc}, improving on the initial fitting by \citet{keownKFPAExaminationsYoung2019} to produce the most robust \nhhh{} analysis of these clouds to date. From this analysis we found that:
\begin{enumerate}
    \item A significant fraction (\percent{13}) of the studied area is best described with a multiple velocity component model. We find that M17, NGC 2264, and W48 have the highest degree of multiple components with $> 15$\%\ of pixels containing two or more velocity structures. M17 and NGC 2264 have the highest fraction (3-4\%) of three-component fits. Conversely W3 West was best fitted with only single velocity components. 
    \item These regions with multiple velocity components tend to be coincident with the densest parts of these clouds, as measured using \nhhh{} Moment 0 or \emph{Herschel} column density maps.  Such areas generally represent the regions of most active star formation in these clouds, suggesting that the multiple velocity components could be connected to hub-filament systems, ridges, or clusters of dense cores. 
    \item When multiple components are included in the analysis, the median measured linewidth of the gas decreases from \kms{0.67} to \kms{0.47} when comparing the multi-component lines with their equivalent single-component widths.  This factor of $\sim$ 1.4 difference could have significant implications for the virial stability of sources in these regions. 
    \item When considering all the clouds together, we find a long tail in the velocity dispersion distribution ($\sigma_\text{obs} > \kms{1}$), which could indicate turbulent-driven feedback.  Nevertheless, when we we consider the clouds separately, that high velocity dispersion tail is mostly dominated by two clouds, M17 and NGC 7538. Broadly, the velocity dispersion varies widely between all the clouds. 
    \item The \ac{ahe} first discussed by \citet{matsakisAnomalousAmmoniaAbsorption1977} is present throughout the high-mass star-forming regions of the \keystone{} survey, which will be a rich dataset for future explorations of this phenomenon.
\end{enumerate}

The resulting data products of our \nhhh{} fitting process are published on \ac{canfar}\footnote{\url{https://www.canfar.net/citation/landing?doi=26.0027}}, and can be used in a variety of future research in high-mass star formation to study gas motions at a range of scales.

\begin{acknowledgements}
TW and SIS acknowledge the support for this work provided by the  \ac{nserc}, RGPIN-2020-03981, RGPIN-2020-03982, and
RGPIN-2020-03983. 
ER acknowledges the support of \ac{nserc}, funding reference number RGPIN-2022-03499.
HK acknowledges support from an \ac{nserc} Discovery Grant.
AG acknowledges support from the  \ac{nsf} under grants AAG 2008101 and CAREER 2142300.
NS acknowledges support by the  CRC1601 (SFB 1601 sub-project B2) funded by the DFG – 500700252.
The National Radio Astronomy Observatory and the Green Bank Observatory are facilities of the National Science Foundation
operated under cooperative agreement by Associated Universities, Inc. 
This research used the \acf{canfar}  operated in partnership by the Canadian Astronomy Data Centre and The Digital Research Alliance of Canada with support from the National Research Council of Canada the Canadian Space Agency, CANARIE and the Canadian Foundation for Innovation.
\end{acknowledgements}

\facilities{GBT}

\software{
This work made use of the following software packages: \formatSoftware{astropy} v7.0.1 \citep{astropy:2013,astropy:2018,astropy:2022}, \formatSoftware{matplotlib} v3.10.3 \citep{Hunter:2007}, \formatSoftware{MUFASA} \href{https://github.com/mcyc/mufasa/tree/82d37a6}{revision 82d37a6} \citep{chenVelocityCoherentFilamentsNGC2020}, \formatSoftware{numpy} \citep{numpy}, \formatSoftware{PySpecKit} v1.0.4 \citep{ginsburgPySpecKitPythonSpectroscopic2011,ginsburgPySpecKitPythonSpectroscopic2022}, \formatSoftware{python} v3.12 \citep{python}, and \formatSoftware{scipy} \citep{2020SciPy-NMeth,scipy14880408}.
Software citation information aggregated in part using \formatSoftware{\href{https://www.tomwagg.com/software-citation-station/}{The Software Citation Station}} \citep{software-citation-station-paper,software-citation-station-zenodo}.
}

\bibliography{QSTAR} 
\bibliographystyle{aasjournalv7}

\appendix
\restartappendixnumbering
\input{Appendices-1_Subclouds}

\input{Appendices-2_TrifitParlims}
\input{Appendices-4_Validation}
\input{Appendices-3_MiscellaneousMaps}

\end{document}

%% file: preamble.tex
\usepackage{acronym}
\usepackage{amsmath, amssymb, amsfonts}
\usepackage{enumitem}
\usepackage{multirow}
\usepackage{tikz}
\usepackage{ulem}
\usepackage{dcolumn}
\usepackage{subfig}
\usepackage{soul}
\usepackage{ulem}

\renewcommand{\toprule}{\hline \hline}
\newcommand{\midrule}{\hline}
\newcommand{\bottomrule}{\hline \hline}

\newcommand{\formatSoftware}[1]{\texttt{#1}}

\acrodefindefinite{GAS}{}{the}
\acrodef{ACORNS}[\formatSoftware{ACORNS}]{Agglomerative Clustering for ORganising Nested Structures}
\acrodef{ahe}[AHE]{anomalous hyperfine emission}
\acrodef{cmb}[CMB]{cosmic microwave background}
\acrodef{cmf}[CMF]{core mass function}
\acrodef{ecr}[ECR]{error-constraining refit}
\acrodef{gas}[GAS]{Green Bank Ammonia Survey}
\acrodef{gbt}[GBT]{Green Bank Telescope}
\acrodef{gmc}[GMC]{giant molecular cloud}
\acrodef{hia}[HIA]{hyperfine intensity anomaly}
\acrodef{hobys}[HOBYS]{\textit{Herschel} OB Young Stars}
\acrodef{iqr}[IQR]{inter-quartile range}
\acrodef{ism}[ISM]{interstellar medium}
\acrodef{KEYSTONE}{KFPA Examination of Young STellar Objects Natal Environments}
\acrodef{kfpa}[KFPA]{K-band Focal Plane Array}
\acrodef{lte}[LTE]{local thermodynamic equilibrium}
\acrodef{mcmc}[MCMC]{Markov chain Monte Carlo}
\acrodef{MUFASA}[\formatSoftware{MUFASA}]{MUlti-component Fitter for Astrophysical Spectral Applications}
\acrodef{rsr}[RSR]{reduced \ac{SCAR} refit}
\acrodef{SCAR}[\formatSoftware{SCAR+}]{Single Component Ammonia Reduction + \ac{MUFASA}}
\acrodef{sed}[SED]{spectral energy distribution}
\acrodef{snr}[S/N]{signal-to-noise ratio}
\acrodef{yso}[YSO]{young stellar object}
\acrodef{aicc}[AICc]{corrected Akaike Information Criterion}
\acrodef{nserc}[NSERC]{Natural Science and Engineering Research Council of Canada}
\acrodef{canfar}[CANFAR]{Canadian Advanced Network For Astronomy Research}
\acrodef{nsf}[NSF]{National Science Foundation}

\newcommand{\keystone}{\ac{KEYSTONE}}
\newcommand{\mufasa}{\ac{MUFASA}}
\newcommand{\pyspeckit}{\formatSoftware{PySpecKit}}
\newcommand{\scarplus}{\ac{SCAR}}

\newcommand{\hz}[2][]{#2~\mbox{#1Hz}} 
\newcommand{\kms}[1]{#1~\mbox{km~s$^{-1}$}}
\newcommand{\pc}[2][]{#2~\mbox{#1pc}}

\newcommand{\K}[1]{#1~\mbox{K}}
\newcommand{\cden}{\mbox{cm$^{-2}$}}

\newcommand{\abs}[1]{\left| #1 \right|}
\newcommand{\percent}[1]{#1\mbox{\%}}

\newcommand{\kb}{k_\text{B}}

\newcommand{\momzero}{Moment 0} 

\newcommand{\tcmb}{T_\text{CMB}}

\newcommand{\tex}{T_\text{ex}}
\newcommand{\tmb}{T_\text{MB}}

\newcommand{\vdobssub}[1]{\sigma_{\text{obs},#1}}
\newcommand{\vdobstext}[1]{\sigma_{\text{obs,#1}}}
\newcommand{\vdobs}{\sigma_\text{obs}}
\newcommand{\vlsrsub}[1]{v_{\text{LSR,}#1}}
\newcommand{\vlsrtext}[1]{v_{\text{LSR,#1}}}
\newcommand{\vlsr}{v_{\text{LSR}}}

\newcommand\nhhh{NH$_3$}
\newcommand\oneone{(1,1)}
\newcommand{\hh}{H$_2$}
\newcommand{\NHH}{\mbox{N(H$_2$)}}

\defcitealias{keownKFPAExaminationsYoung2019}{K19}

\newcommand{\kt}{\citetalias{keownKFPAExaminationsYoung2019}}
\newcommand{\inprep}[1]{(#1 et al. in preparation)}

\newcommand{\refapp}[1]{Appendix~\ref{#1}}
\newcommand{\refsec}[1]{Section~\ref{#1}}
\newcommand{\refchap}[1]{\refsec{#1}}
\newcommand{\refeq}[1]{Equation~\ref{#1}}
\newcommand{\reffig}[1]{Figure~\ref{#1}}
\newcommand{\reftab}[1]{Table~\ref{#1}}

\definecolor{myblue}{RGB}{0,0,220}
\definecolor{mygreen}{RGB}{0,128,0}
\definecolor{mypurple}{RGB}{138,43,226}
\definecolor{myred}{RGB}{220,20,60}

\newcommand{\rnote}[1]{\footnote{}}

\newcommand{\sis}[1]{} 
\newcommand{\jdf}[1]{} 
\newcommand{\rf}[1]{} 

\newcolumntype{.}{D{.}{.}{-1}} 
\newcolumntype{i}{D{.}{.}{0}} 

\newcommand{\tallth}[1]{\th{\multirow{2}{*}{#1}}}
\renewcommand{\th}[1]{\multicolumn{1}{c}{\text{#1}}}

\setlist[1]{leftmargin=2\parindent} 
\setlist{nosep}

\usetikzlibrary{shapes, arrows, positioning}
\tikzset{node/.style={text centered, minimum height=1.75em, minimum width=4em, draw=black, align=center }}
\tikzstyle{arrow}=[thick,->]
\tikzstyle{data}=[trapezium, trapezium left angle=80, trapezium right angle=100, fill=blue!30,inner xsep=.3em,style=node]
\tikzstyle{decision}=[diamond, aspect=3, fill=green!30, inner sep=.2em, style=node]
\tikzstyle{option}=[circle, text centered, draw=black, fill=white]
\tikzstyle{process}=[rectangle, fill=orange!30, style=node]
\tikzstyle{terminator}=[rectangle, rounded corners, fill=red!30, style=node]

%% file: Tables-TargetClouds.tex
\begin{tabular}{lrriD{/}{\pm}{-1}cD{.}{.}{3}} \toprule
\tallth{Region} & \th{R.A.}    & \th{Dec.}    & \th{Area}         & \th{Distance} & \tallth{Ref.} & \th{rms} \\
& \th{(J2000)} & \th{(J2000)} & \th{(arcmin$^2$)} & \th{(kpc)}    &                & \th{(K)} \\ \midrule
Cygnus X North   & 20h38\text{m}12.2\text{s} & $42\text{:}02\text{:}05$  & 4747 & 1.35/0.1  & (1) & 0.095 \\
Cygnus X South   & 20h33\text{m}32.5\text{s} & $39\text{:}28\text{:}58$  & 5264 & 1.49/0.1  & (1) & 0.094 \\
M16              & 18h18\text{m}27.6\text{s} & $-13\text{:}42\text{:}00$ & 681  & 1.69/0.01 & (2) & 0.116 \\
M17              & 18h20\text{m}01.1\text{s} & $-16\text{:}12\text{:}55$ & 1019 & 1.52/0.08 & (2) & 0.116 \\
MonR1            & 6h32\text{m}32.3\text{s}  & $10\text{:}22\text{:}12$  & 715  & 0.72/0.05 & (3) & 0.092 \\
MonR2            & 6h07\text{m}56.6\text{s}  & $-6\text{:}18\text{:}36$  & 751  & 0.86/0.02 & (2) & 0.096 \\
NGC 2264         & 6h41\text{m}12.9\text{s}  & $9\text{:}33\text{:}29$  & 1231 & 0.75/0.02 & (2) & 0.103 \\
NGC 7538         & 23h14\text{m}47.9\text{s} & $61\text{:}28\text{:}47$  & 2193 & 2.65/0.1  & (4) & 0.093 \\
Rosette          & 6h33\text{m}24.7\text{s}  & $4\text{:}25\text{:}53$   & 1979 & 1.49/0.04 & (5) & 0.095 \\
W3 Main\textsuperscript{a} & 2h23\text{m}37.4\text{s}  & $61\text{:}44\text{:}03$  & 2802 & 1.91/0.11 & (2) & 0.092 \\
W3 West\textsuperscript{a} & 2h21\text{m}03.6\text{s}  & $61\text{:}14\text{:}39$  & 927  & 1.91/0.11 & (2) & 0.107 \\
W48              & 18h59\text{m}34.0\text{s} & $1\text{:}22\text{:}21$   & 2022 & 2.3/0.15  & (6) & 0.100   \\
\bottomrule \end{tabular}

%% file: Figures-MomentZeroMaps_Example.tex
\begin{figure*}[htb!]
    \centering
    \includegraphics[width=0.9\linewidth]{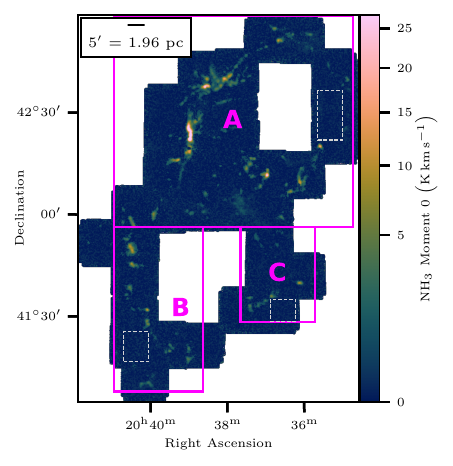}
    \caption{\label{fig:moment_zero_with_overlays}Cygnus X North: \momzero{} map using data from \kt. Magenta boxes show the subregions used for our analysis. Grey boxes show the empty regions used for noise characterization. Areas of low general emission (such as on the edges or left hand side) are in some cases excluded.}
\end{figure*}

%% file: Figures-Flowchart3cFitting.tex
\begin{tikzpicture}
\tikzstyle{steps1}=[text width=1.5in]
\tikzstyle{steps2}=[text width=2in]

\node (start) [terminator] {Start};

\node (errfit) [process,below=1em of start,steps1] {0. Refit to constrain errors};
\draw [arrow] (start)--(errfit);

\node (mask) [process,below=1em of errfit,steps1] {1. Mask the data};
\draw [arrow] (errfit)--(mask);

\node (MUFASA) [process,below=1em of mask,steps1] {2. Run \acs{MUFASA}};
\draw [arrow] (mask) -- (MUFASA);

\node (mask2) [process,below=1em of MUFASA,steps1] {3. Create a new mask};
\draw [arrow] (MUFASA) -- (mask2);

\node (gridding) [process,below=1em of mask2,steps1] {4. Divide the data into subgrids};
\draw [arrow] (mask2)--(gridding);

\node (forloop) [rectangle, rounded corners, fill=black!20, align=left,below=1em of gridding.south, minimum width=3.352in, minimum height=3.9in, anchor=north] {};
\node () [rectangle, below right=.2em of forloop.north west] {For each subgrid};

\node (anchor) [below = .5em of forloop.north] {};
\node (gridcount) [decision,right=3.5em of anchor, anchor=north] {How many pixels have fits?};
\node (joinerA) [rectangle, above=1em of gridcount] {};
\draw [arrow] (gridding.south)|-(joinerA.center)--(gridcount.north);

\node (under10) [option, left=1em of gridcount, anchor=east, font=\footnotesize] {$<10$};
\draw [arrow] (gridcount)-- (under10);

\node (over10) [option, below=1em of gridcount, font=\footnotesize] {$\geq10$};
\draw [arrow] (gridcount)-- (over10);

\node (output2c) [data,below=2.25em of under10,text width=1.25in] {Output 2-component parameters};
\node (joinerB) [rectangle, above=1em of output2c] {};
\draw [arrow] (under10.south)|-(joinerB.center)--(output2c.north);

\node (findbright) [process,steps2,below=1em of over10] {5. Identify the brightest component in the subgrid};
\draw [arrow] (over10)--(findbright);

\node (fitbright) [process,steps2,below=1em of findbright] {6. Fit the bright component over the entire subgrid, and subtract this model from the grid};
\draw [arrow] (findbright)--(fitbright);

\node (fitresids) [process,steps2,below=1em of fitbright] {7. Fit the residual cube with \acs{MUFASA}};
\draw [arrow] (fitbright)--(fitresids);

\node (fit3c) [process,steps2,below=1em of fitresids] {8. Fit bright and \acs{MUFASA} components simultaneously};
\draw [arrow] (fitresids)--(fit3c);

\node (output3c) [data,below=1em of fit3c,text width =1.5in] {Output 3-component parameters};
\draw [arrow] (fit3c)--(output3c);

\node (regrid) [data,below=1em of output3c,text width=1.5in] {Recombine subgrid parameter cubes};
\draw [arrow] (output3c)--(regrid);
\draw [arrow] (output2c)|-(regrid);

\node (end) [terminator, below=1em of regrid] {End};
\draw [arrow] (regrid)--(end);
\end{tikzpicture}

%% file: Tables-OverlapPotentialValues.tex
\begin{tabular}{l....} \toprule
\tallth{Row} & \tallth{$\Delta v$} & \tallth{$\Delta v-\ell$} & \tallth{$\Delta v-r$} & \tallth{SP}\\[3mm]
& \multicolumn{4}{c}{$\left(\mathrm{km\,s^{-1}}\right)$} \\ \midrule
Top    & 3.00 & 4.47  & 10.57 & 3.00 \\
Middle & 0.50 & 6.97  & 8.07  & 0.50 \\
Bottom & 7.00 & 14.47 & 0.57  & 0.57 \\
\bottomrule \end{tabular}

%% file: Figures-FlowchartDecisionTree.tex
\tikzstyle{tw}=[text width=1.5in, minimum height=4em, aspect=2, inner sep=.2em]
\begin{tikzpicture}

\node (start) [terminator] {Start};

\node (1) [decision, tw, below=1em of start] {Is the refitted model statistically preferred ($\ln(K)>0$)?};
\node (2) [decision, tw, below=1.5em of 1] {Is the separation in velocities\\large (\kms{\textgreater~0.28})?};
\node (3) [decision, tw, below=1.5em of 2] {Is the primary model strongly preferred ($\ln(K) <-50$)?};
\node (4) [decision, tw, below=1.5em of 3] {Do the kinematics change significantly?};

\node (original) [terminator, below left=3em of 4.south, text width =1.2in, inner xsep=.2em] {Choose original model.};
\node (refit) [terminator,below right=3em of 4.south, text width =1.2in, inner xsep=.2em] {Choose \ac{rsr} model.};

\node (refitTop) [rectangle, above=2em of refit.east] {};
\node (originalTop) [rectangle, above=2em of original.west] {};

\draw [arrow] (start)--(1);
\draw [arrow] (1)-- node[anchor=east] {No} (2);
\draw [arrow] (1)-| node[anchor=south east] {Yes} (refitTop.center) -| (refit.north);
\draw [arrow] (2)-- node[anchor=east] {No} (3);
\draw [arrow] (2)-| node[anchor=south west] {Yes} (originalTop.center) -| (original.north);
\draw [arrow] (3)-- node[anchor=east] {No} (4);
\draw [arrow] (3)-| node[anchor=south west] {Yes} (originalTop.center) -| (original.north);
\draw [arrow] (4)-| node[anchor=south west] {Yes} (originalTop.center) -| (original.north);
\draw [arrow] (4)-| node[anchor=south east] {No} (refitTop.center) -| (refit.north);
\end{tikzpicture}

%% file: Tables-ComponentStatistics.tex
\begin{tabular}{li...} \toprule
\tallth{Region} & \tallth{Fitted Pixels} & \multicolumn{3}{c}{Components (\%)}\\
& & \th{1} & \th{2} & \th{3}\\ \midrule
Cygnus X North & 10533 & 89.1  & 9.8  & 1.1 \\
Cygnus X South & 4483  & 91.2  & 7.3  & 1.5 \\
M16            & 3218  & 85.6  & 13.0 & 1.4 \\
M17            & 3177  & 73.2  & 22.2 & 4.6 \\
MonR1          & 2236  & 86.7  & 12.4 & 0.9 \\
MonR2          & 3319  & 96.8  & 3.2  & 0.0 \\
NGC 2264       & 4504  & 76.6  & 20.2 & 3.2 \\
NGC 7538       & 4632  & 89.2  & 10.3 & 0.5 \\
Rosette        & 1874  & 99.1  & 0.9  & 0.0 \\
W3 Main        & 2929  & 90.0  & 9.7  & 0.4 \\
W3 West        & 368   & 100.0 & 0.0  & 0.0 \\
W48            & 8727  & 81.5  & 16.0 & 2.5 \\ \midrule
All            & 50000 & 86.5  & 11.9 & 1.6 \\
\bottomrule \end{tabular}

%% file: Tables-ComponentSNRs.tex
\begin{tabular}{l...} \toprule
\tallth{Region}  & \multicolumn{3}{c}{Number of Components}\\
& \th{1} & \th{2} & \th{3}\\ \midrule
Cygnus X North & 5.59 & 17.22    & 26.86    \\
Cygnus X South & 5.72 & 16.85    & 23.2     \\
M16            & 5.49 & 15.29    & 29.26    \\
M17            & 5.29 & 16.66    & 27.16    \\
MonR1          & 5.41 & 13.81    & 20.95    \\
MonR2          & 5.43 & 14.13    & 9.18     \\
NGC 2264       & 5.4  & 16.59    & 22.55    \\
NGC 7538       & 5.12 & 13.73    & 12.15    \\
Rosette        & 4.91 & 12.36    & \th{N/A} \\
W3 Main        & 5.53 & 12.86    & 12.33    \\
W3 West        & 4.86 & \th{N/A} & \th{N/A} \\
W48            & 5.47 & 15.23    & 24.15    \\
\bottomrule \end{tabular}

%% file: Tables-ModelStatistics.tex
\begin{tabular}{l...} \toprule
\tallth{Model} & \multicolumn{3}{c}{Components (\%)}\\
& \th{1} & \th{2} & \th{3}\\ \midrule
\ac{ecr}    & 82.1 & \th{N/A} & \th{N/A} \\
\mufasa{} 1 & 1.2  & \th{N/A} & \th{N/A} \\
\mufasa{} 2 & 0.1  & 10.4     & \th{N/A} \\
\scarplus   & 1.0  & 1.4      & 1.6      \\
\ac{rsr}    & 2.0  & 0.1      & \th{N/A} \\ \midrule
Total       & 86.5 & 11.9     & 1.6      \\
\bottomrule \end{tabular}

%% file: Tables-ComponentPercentilesMom0.tex
\begin{tabular}{cc...ccccc} \toprule
\multicolumn{2}{c}{\momzero{} Bins} & \multicolumn{3}{c}{Components (\%)} & \multicolumn{3}{c}{Median Component S/N} & \multicolumn{2}{c}{$\vdobs\ \left(\mathrm{km\,s^{-1}}\right)$\tablenotemark{a}}\\
\th{Percentiles} & \th{Values}  & \th{1} & \th{2} & \th{3} & \th{1} & \th{2} & \th{3} & \th{\ac{ecr}} & \th{\ac{SCAR}} \\ \midrule

0---10   & 0.20---0.67  & 100.0 & 0.0  & 0.0  & $3.7\substack{+0.8\\-0.5}$  & \mbox{$\cdots$}              & \mbox{$\cdots$}               & $0.27\substack{+0.13\\-0.11}$ & $0.27\substack{+0.13\\-0.11}$ \\
10---20  & 0.67---0.87  & 100.0 & 0.0  & 0.0  & $4.0\substack{+1.2\\-0.7}$  & \mbox{$\cdots$}              & \mbox{$\cdots$}               & $0.35\substack{+0.15\\-0.14}$ & $0.35\substack{+0.15\\-0.14}$ \\
20---30  & 0.87---1.08  & 100.0 & 0.0  & 0.0  & $4.4\substack{+1.5\\-1.0}$  & $8.5\substack{+0.0\\-0.0}$   & \mbox{$\cdots$}               & $0.40\substack{+0.18\\-0.15}$ & $0.40\substack{+0.18\\-0.15}$ \\
30---40  & 1.08---1.32  & 99.9  & 0.1  & 0.0  & $4.8\substack{+1.8\\-1.2}$  & $8.5\substack{+0.6\\-0.6}$   & \mbox{$\cdots$}               & $0.44\substack{+0.23\\-0.17}$ & $0.44\substack{+0.23\\-0.17}$ \\
40---50  & 1.32---1.63  & 99.4  & 0.6  & 0.0  & $5.4\substack{+2.2\\-1.6}$  & $7.5\substack{+0.6\\-0.4}$   & \mbox{$\cdots$}               & $0.48\substack{+0.26\\-0.19}$ & $0.47\substack{+0.26\\-0.19}$ \\
50---60  & 1.63---2.06  & 97.5  & 2.5  & 0.0  & $6.4\substack{+2.6\\-2.1}$  & $8.6\substack{+1.5\\-1.1}$   & \mbox{$\cdots$}               & $0.48\substack{+0.29\\-0.19}$ & $0.47\substack{+0.30\\-0.19}$ \\
60---70  & 2.06---2.72  & 92.7  & 7.2  & 0.1  & $7.8\substack{+3.3\\-2.7}$  & $9.0\substack{+2.5\\-1.5}$   & $9.1\substack{+0.1\\-0.3}$    & $0.49\substack{+0.33\\-0.19}$ & $0.46\substack{+0.33\\-0.19}$ \\
70---80  & 2.72---3.92  & 81.0  & 18.5 & 0.5  & $10.0\substack{+4.5\\-3.4}$ & $10.7\substack{+3.2\\-2.3}$  & $10.1\substack{+6.6\\-1.4}$   & $0.50\substack{+0.34\\-0.18}$ & $0.44\substack{+0.33\\-0.17}$ \\
80---90  & 3.92---6.60  & 59.0  & 38.7 & 2.2  & $14.0\substack{+5.4\\-4.5}$ & $13.9\substack{+4.4\\-3.6}$  & $14.2\substack{+4.5\\-4.1}$   & $0.54\substack{+0.31\\-0.18}$ & $0.47\substack{+0.32\\-0.19}$ \\
90---100 & 6.60---54.21 & 35.4  & 51.4 & 13.1 & $21.8\substack{+8.6\\-6.8}$ & $23.3\substack{+13.9\\-6.8}$ & $27.9\substack{+15.5\\-10.3}$ & $0.69\substack{+0.43\\-0.22}$ & $0.57\substack{+0.38\\-0.23}$ \\
\bottomrule \end{tabular}

%% file: Tables-ComponentPercentilesNH2.tex
\begin{tabular}{cc...cc} \toprule
\multicolumn{2}{c}{log$_{10}$N(\hh) Bins} & \multicolumn{3}{c}{Components (\%)} & \multicolumn{2}{c}{$\vdobs\ \left(\mathrm{km\,s^{-1}}\right)$\tablenotemark{a}}\\
\th{Percentiles} & \th{Values} & \th{1} & \th{2} & \th{3} & \th{\ac{ecr}} & \th{\ac{SCAR}} \\ \midrule
0---10   & 21.44---22.06 & 99.9 & 0.1  & 0.0  & $0.27\substack{+0.17\\-0.12}$ & $0.27\substack{+0.17\\-0.12}$ \\
10---20  & 22.06---22.14 & 99.6 & 0.4  & 0.0  & $0.34\substack{+0.19\\-0.14}$ & $0.34\substack{+0.19\\-0.14}$ \\
20---30  & 22.14---22.20 & 98.9 & 1.1  & 0.0  & $0.36\substack{+0.19\\-0.14}$ & $0.36\substack{+0.20\\-0.14}$ \\
30---40  & 22.20---22.25 & 98.3 & 1.7  & 0.0  & $0.39\substack{+0.19\\-0.14}$ & $0.39\substack{+0.20\\-0.14}$ \\
40---50  & 22.25---22.30 & 97.1 & 2.9  & 0.0  & $0.43\substack{+0.22\\-0.14}$ & $0.42\substack{+0.22\\-0.15}$ \\
50---60  & 22.30---22.35 & 94.1 & 5.8  & 0.1  & $0.47\substack{+0.24\\-0.15}$ & $0.45\substack{+0.25\\-0.16}$ \\
60---70  & 22.35---22.42 & 88.3 & 11.4 & 0.3  & $0.50\substack{+0.26\\-0.16}$ & $0.48\substack{+0.27\\-0.18}$ \\
70---80  & 22.42---22.51 & 79.8 & 19.1 & 1.1  & $0.53\substack{+0.28\\-0.17}$ & $0.48\substack{+0.31\\-0.18}$ \\
80---90  & 22.51---22.66 & 67.5 & 30.0 & 2.5  & $0.60\substack{+0.28\\-0.20}$ & $0.52\substack{+0.30\\-0.21}$ \\
90---100 & 22.66---23.63 & 41.7 & 46.5 & 11.8 & $0.77\substack{+0.47\\-0.24}$ & $0.62\substack{+0.40\\-0.26}$ \\
\bottomrule \end{tabular}

%% file: Appendices-1_Subclouds.tex
\section{(Sub)cloud Details}\label{app:subclouds}

This Appendix details how the large \keystone{} regions were subdivided.
Figure \ref{fig:app_mom0} shows the \momzero{} maps as generated by \kt{} in alphabetical order, with Cygnus X North shown in Figure \ref{fig:moment_zero_with_overlays}.
Cloud subregions are identified using pink boxes and labels, and the blank regions used to measure the noise are shown by the thin white boxes.
\reftab{tab:SubcloudCoordinates} gives the pixel ranges used to define the boundaries of the subclouds (based on the pixel dimensions from \kt) for the seven KEYSTONE clouds that were broken up for analysis.
\reftab{tab:Mom0NoiseRegions} lists the pixel ranges for each cloud denoting the areas used to calculate the \momzero{} noise.

\begin{table}[htb!]\centering
    \caption{\label{tab:SubcloudCoordinates}Subcloud Boundaries}
    \input{Tables-SubcloudCoordinates}
\end{table}

\begin{table}[htb!]\centering
	\caption{\label{tab:Mom0NoiseRegions}\momzero{} Noise Areas}
    \input{Tables-Mom0NoiseRegions}
\end{table}

\input{Figures-MomentZeroMaps_All}

%% file: Tables-SubcloudCoordinates.tex
\begin{tabular}{lD{-}{-}{3.3}D{-}{-}{3.3}} \toprule
\th{Subcloud Name} &    \th{X (pixels)}    &    \th{Y (pixels)}    \\ \midrule
Cygnus X North A & 70-550  & 350-775 \\
Cygnus X North B & 70-250  & 20-350  \\
Cygnus X North C & 325-475 & 160-350 \\
Cygnus X South A & 433-700 & 625-824 \\
Cygnus X South B & 30-450  & 340-630 \\
Cygnus X South C & 225-750 & 5-250   \\
M17 A            & 20-150  & 110-325 \\
M17 B            & 150-250 & 0-150   \\
NGC 7538 East    & 50-200  & 10-350  \\
NGC 7538 West    & 200-360 & 100-225 \\
Rosette A        & 75-300  & 70-400  \\
Rosette B        & 375-475 & 130-210 \\
W3 Main A        & 90-425  & 300-525 \\
W3 Main B        & 485-583 & 323-420 \\
W3 Main C        & 0-230   & 0-275   \\
W48 A            & 400-500 & 200-370 \\
W48 B            & 475-630 & 65-225  \\
W48 C            & 10-390  & 0-200   \\
\bottomrule  \end{tabular}

%% file: Tables-Mom0NoiseRegions.tex
\begin{tabular}{lD{-}{-}{3.3}D{-}{-}{3.3}} \toprule
\th{(Sub)cloud Name} &    \th{X (pixels)}    &    \th{Y (pixels)}    \\ \midrule
Cygnus X North A & 410-460 & 175-275 \\
Cygnus X North B & 20-70   & 60-120  \\
Cygnus X North C & 60-110  & 0-45    \\
Cygnus X South A & 160-210 & 0-50    \\
Cygnus X South B & 210-260 & 150-230 \\
Cygnus X South C & 50-220  & 100-150 \\
M16              & 180-205 & 65-100  \\
M17 A            & 0-40    & 70-150  \\
M17 B            & 0-60    & 70-110  \\
MonR1            & 100-235 & 190-216 \\
MonR2            & 85-135  & 110-145 \\
NGC 2264         & 190-240 & 250-290 \\
NGC 7538 East    & 100-150 & 200-270 \\
NGC 7538 West    & 80-150  & 90-120  \\
Rosette A        & 75-150  & 0-60    \\
Rosette B        & 50-100  & 0-50    \\
W3 Main A        & 120-170 & 0-110   \\
W3 Main B        & 10-80   & 50-80   \\
W3 Main C        & 100-175 & 75-150  \\
W3 West          & 120-170 & 100-140 \\
W48 A            & 65-100  & 0-60    \\
W48 B            & 90-150  & 10-70   \\
W48 C            & 220-320 & 75-150  \\
\bottomrule  \end{tabular}

%% file: Figures-MomentZeroMaps_All.tex
\begin{figure*}[htb!]
\centering
    \includegraphics[width=0.49\linewidth]{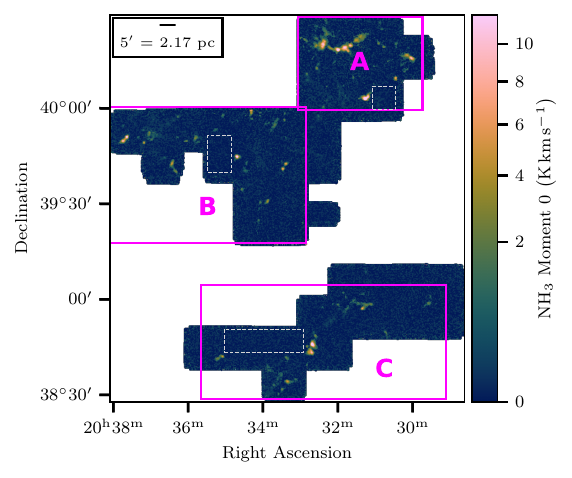} \includegraphics[width=0.49\linewidth]{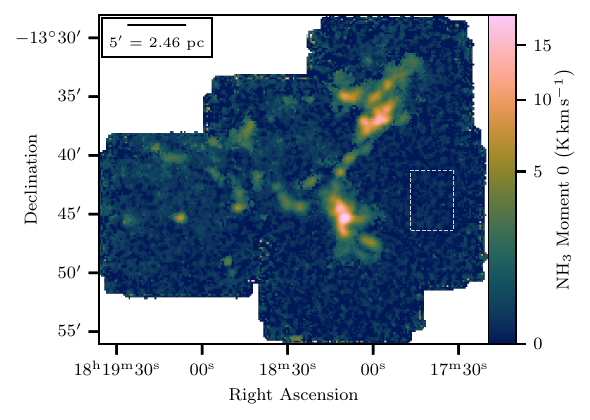} 
    \qquad
    \includegraphics[width=0.49\linewidth]{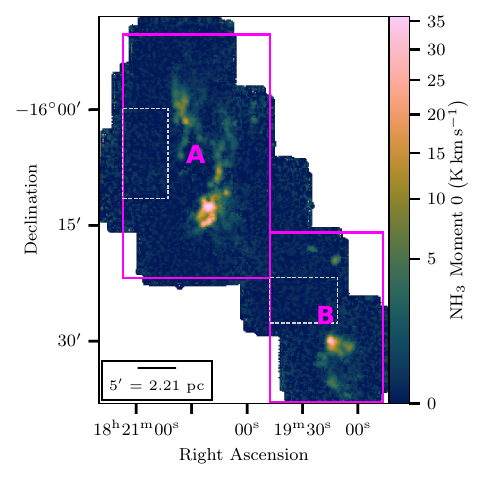} 
    \includegraphics[width=0.49\linewidth]{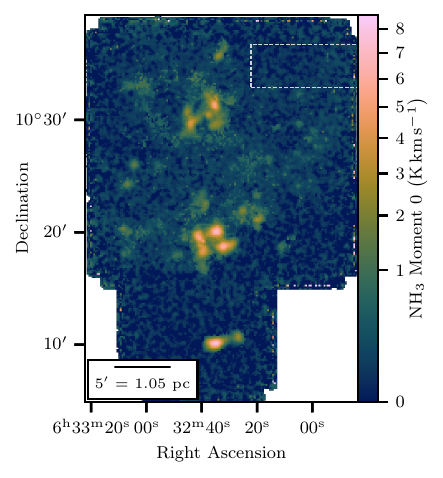}
    \caption{Moment 0 maps for the \keystone{} clouds.  Same as \reffig{fig:moment_zero_with_overlays} for Cygnus X South (upper-left) and M16 (upper-right), M17 (lower-left) and Mon R1 (lower-right).\label{fig:app_mom0}}
\end{figure*}

\begin{figure*}[htb!]
    \centering
    \ContinuedFloat
    \includegraphics[width=0.49\linewidth]{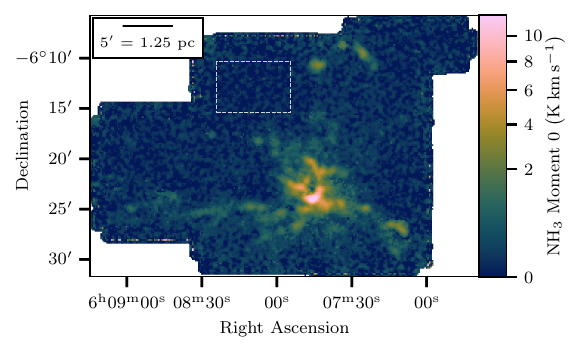}
    \includegraphics[width=0.49\linewidth]{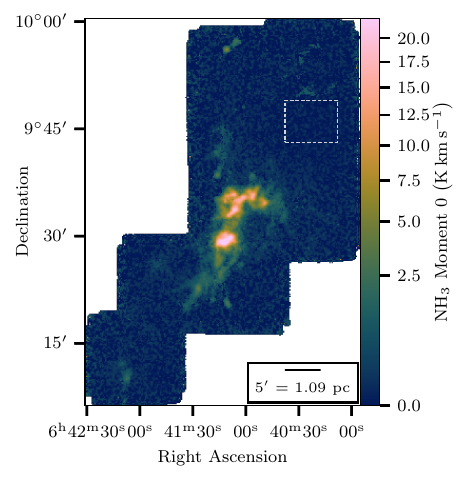}
    \qquad
    \includegraphics[width=0.49\linewidth]{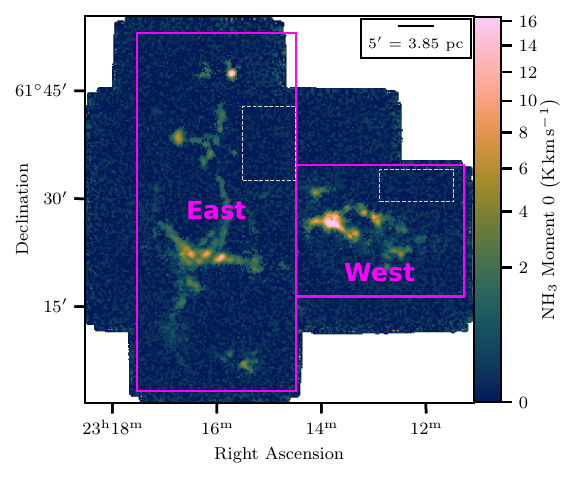}
    \includegraphics[width=0.49\linewidth]{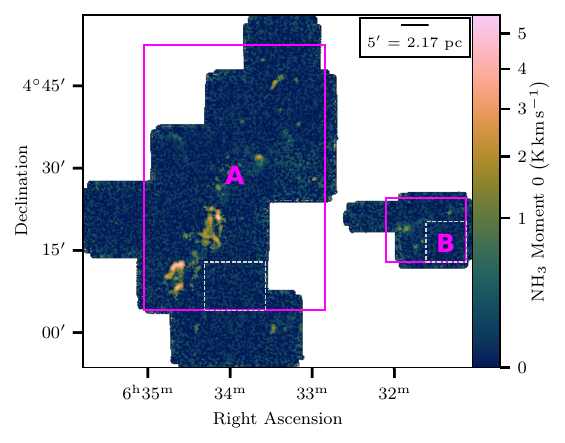}
    \caption{Continued - for Mon R2 (upper-left), NGC 2264 (upper-right), NGC 7538 (lower-left), Rosette (lower-right).}
\end{figure*}

\begin{figure*}[htb!]
    \centering
    \ContinuedFloat
    \includegraphics[width=0.49\linewidth]{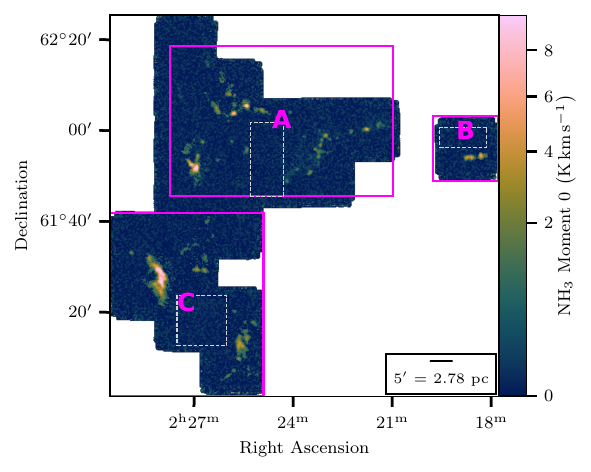}
	\includegraphics[width=0.49\linewidth]{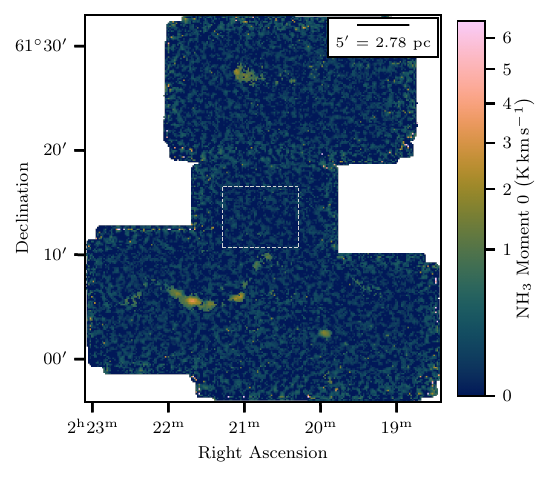}
    \qquad
    \includegraphics[width=0.65\linewidth]{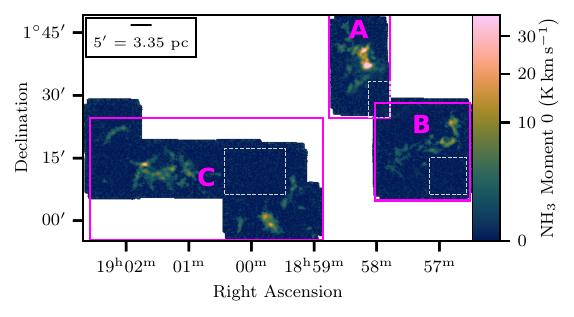}
    \caption{Continued - for W3 Main (upper-left), W3 West (upper-right), W48 (lower).}
\end{figure*}

%% file: Appendices-2_TrifitParlims.tex
\section{Three-Component Kinematic Parameter Limits}\label{app:trifitparlims}
For Step 8 of our \scarplus{} routine (see \reffig{fig:Flowchart3cFitting} and \refsec{sec:ThreeComponentFitting}), we fit three components simultaneously over a single subgrid of $5\times5$ pixels with the goal of constraining the excitation temperature and optical depth via radiative transfer.
Since the kinematic parameters have been constrained by the previous steps, we expect minimal change in the centroid velocity and velocity dispersion during this fit.

Our fitting model is based on the \pyspeckit{} package and makes use of its ability to run in parallel when fitting over a cube.
While fitting each pixel individually would allow for fine-tuned control over the parameter limits for each pixel, when fitting a cube \pyspeckit{}, however, only accepts one global set of parameter limits which are applied to the entire cube.
Thus for Step 8 we must generate twelve sets of minimum and maximum parameter values --- corresponding to the four parameters ($\vlsr$, $\vdobs$, $\tex$, $\tau$) in each of the three velocity components.
These twelve sets of parameter limits will be applied over the entire subgrid, as required by \pyspeckit.

Since the kinematic parameters are well constrained by the previous steps, the Step 8 fit is intended to constrain better the excitation temperature and optical depth.
Therefore, $\tex$ and $\tau$ are free parameters for all three components and we use restrictive parameter limits for the centroid velocity and velocity dispersion according to the previously obtained three-component model.
\reftab{tab:trifit_parlims} provides a summary of how these parameter limits are generated.

\begin{table}[htb!]{\centering
	\caption{\label{tab:trifit_parlims} Definition of Parameter Limits for \acs{SCAR} Step~8}

    \begin{tabular}{lll} \toprule
                & \th{$\vlsr$} & \th{$\vdobs$}\\ \midrule
    Component 1 & Step 6 & Step 6 \\
    Component 2 & \refeq{eq:combine23-vlsr-ranges} & \refeq{eq:combine23-vdisp-ranges} \\
    Component 3 & \refeq{eq:combine23-vlsr-ranges} & \refeq{eq:combine23-vdisp-ranges} \\
    \bottomrule \end{tabular}}
    \tablecomments{The excitation temperature and optical depth limits for all components are taken from \reftab{tab:DefaultParameterLimits}.}
\end{table}

The origin of the each of three components provides guidance on how to choose parameter limits. 
Component one is expected to be continuous over the grid, and was already fit with strict centroid velocity and velocity dispersion limits (Step 6), so these limits will be retained for the Step 8 thee-component re-fit.
Components two and three were identified simultaneously by \ac{MUFASA} (Step 7).
As \ac{MUFASA} has no intrinsic component ordering, a structure that is labelled as component two in one pixel, may instead be labelled as component three in another (see \reffig{fig:ComponentSwitchingExample}).
Because of this ambiguity, and the fact that our parameter limits cannot be modified per-pixel, we create a broad set of limits that encompass both the second and third component and apply this set of limits to both.

\begin{figure*}[htb!]
    \centering
    \includegraphics[width=.75\linewidth]{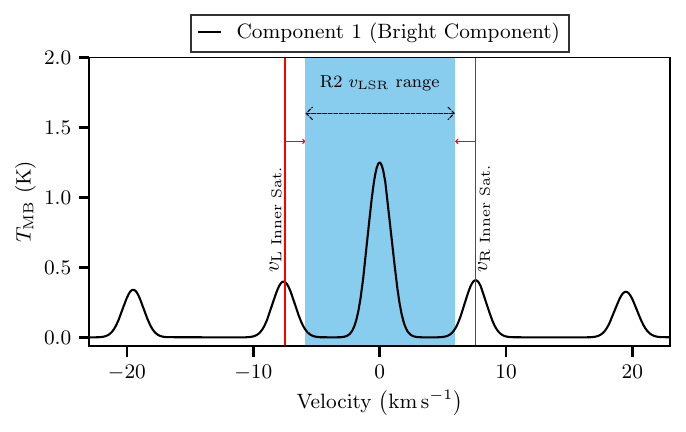}
    \caption{\label{fig:TrifitCentroidLimits}The range of centroid velocities that Requirement 2 (see text) allows for the second and third components. The vertical red lines indicate the locations of the inner satellites for component one, and the red arrows show the restriction of the range by $2\vdobssub{\text{1c}}$ (the fitted velocity dispersion of the bright component, as found in Step 6) on each side. The shaded blue region is the resulting range of potential velocity centroid values.
    }
\end{figure*}

The requirements that we wish to apply to the second and third components are as follows:
\begin{enumerate}
    \item the components identified by \ac{MUFASA} in Step 7 are correct, and therefore the parameters should not be allowed to vary outside the already-identified values; and
    \item the primary hyperfine groups of components two and three should lie within the inner satellite groups of component one.
\end{enumerate}
These are assumptions that we make to create a sufficiently limited parameter space. 
\reffig{fig:TrifitCentroidLimits} shows graphically the meaning of Requirement 2.

To satisfy Requirement 1 (R1), we generate centroid velocity limits from the following equations:
\begin{equation}\label{eq:mufasa23-vlsr-lims}\begin{split}
    \vlsrtext{min,R1} &= \min{(\vlsrtext{2c}, \vlsrtext{3c})}- \kms{0.5}, \\
    \vlsrtext{max,R1} &= \max{(\vlsrtext{2c}, \vlsrtext{3c})}+ \kms{0.5},\\
\end{split}\end{equation}
where $\vlsrtext{2c}$ and $\vlsrtext{3c}$ are the centroid velocity values output by \mufasa{} during Step 7 (fitting the residual subgrid).
In \refeq{eq:mufasa23-vlsr-lims}, the addition of \kms{0.5} (just over three channel widths) is intended to add some additional room to the limits, under the assumption that the \mufasa{} Step 7 parameters may not be exactly accurate, but are unlikely to be significantly off.
We want to ensure, however, that these limits are not overly restrictive, and therefore apply the following logic:

\begin{gather}\label{eq:minimum23-vlsr-range}
    \text{if } (\vlsrtext{max,R1}-\vlsrtext{min,R1})<\kms{2} \nonumber \\
    \text{then }
    \begin{cases}
        \vlsrtext{min,R1} = \Delta \vlsr - \kms{1} \\
        \vlsrtext{max,R1} = \Delta \vlsr + \kms{1} \\
    \end{cases} \\
    \text{where } \Delta \vlsr = \frac{1}{2}(\vlsrtext{max,R1}+\vlsrtext{min,R1}). \nonumber
\end{gather}

To satisfy Requirement 2 (R2), we generate centroid velocity limits from the following equations:
\begin{equation}\label{eq:satellite23-vlsr-range}\begin{split}
    \vlsrtext{min., R2} = v_\text{1c Left Inner Sat.} + 2\vdobssub{\text{1c}}, \\
    \vlsrtext{max., R2} = v_\text{1c Right Inner Sat.} - 2\vdobssub{\text{1c}}.
\end{split}\end{equation}
This range of values is represented in \reffig{fig:TrifitCentroidLimits}.

We now have a set of centroid velocity limits which satisfies Requirement 1, and one which satisfies Requirement 2.
We take the union of these two to generate our final set of limits as follows:
\begin{equation}\label{eq:combine23-vlsr-ranges}\begin{split}
    \vlsrtext{min} &= \max{(\vlsrtext{min,R1}, \vlsrtext{min,R2})} \\
    \vlsrtext{max} &= \min{(\vlsrtext{max,R1}, \vlsrtext{max,R2})}
\end{split}\end{equation}
This equation typically results in a set of limits more restrictive than is required by either Requirement 1 or Requirement 2, but that follows both.

For the velocity dispersion, we follow a method similar to that used for the centroid velocity.
\begin{equation}\label{eq:mufasa23-vdisp-range}\begin{split}
    \vdobstext{min,R1} &= \min{(\vdobstext{2c}, \vdobstext{3c})}- \kms{0.25}, \\
    \vdobstext{max,R1} &= \max{(\vdobstext{2c}, \vdobstext{3c})}+ \kms{0.25},
\end{split}\end{equation}
\begin{equation}\label{eq:combine23-vdisp-ranges}\begin{split}
    \vdobstext{min} &= \max{(\vdobstext{min,R1}, \kms{0.14})}, \\
    \vdobstext{max} &= \min{(\vdobstext{max,R1}, \kms{1.75})},
\end{split}\end{equation}
\refeq{eq:mufasa23-vdisp-range} is analogous to \refeq{eq:mufasa23-vlsr-lims}, where here
where $\vdobstext{2c}$ and $\vdobstext{3c}$ are the velocity dispersions, and we add extra space for the fitter of \kms{0.25} (reduced from \kms{0.5}). 
Here we add a smaller amount of additional parameter space, following the smaller range of values that velocity dispersions can take.
In \refeq{eq:combine23-vdisp-ranges}, we apply the $\vdobs$ restrictions from \refeq{eq:mufasa23-vdisp-range} in concert with the global restrictions given in \reftab{tab:DefaultParameterLimits}.

%% file: Appendices-4_Validation.tex
\section{Validation of the \scarplus{} Algorithm}\label{app:validation}

Here we details the steps taken to validate the output of the \scarplus{} algorithm.
Specifically, we wish to test that given a mock spectra with a known number of velocity components with known parameters, these velocity components are accurately identified by \scarplus.
We note that validation of the \mufasa{} algorithm, verifying that it accurately reproduces the intended number of components when run on mock spectra, was previously performed by \citet{chenVelocityCoherentFilamentsNGC2020}.

\subsection{Creation of Mock Data}

Because \scarplus{} applies model fitting over maps of data which are expected to have some level of continuity rather than single spectral pixels, we require that any mock data used for validation also has this property.
Nevertheless, it is a significant undertaking to generate large-scale smoothly varying parameter maps where the number of components is variable as well as ensuring those parameter maps encompass the variety of subtleties present in the \keystone{} data. 

Therefore, in lieu of true mock data, we utilize the best-fit parameters from one of the \keystone{} \acp{gmc}.
For this, we chose NGC2264, as it is primarily a single contiguous region, and contains high levels of multi-component pixels.
To create a cube of mock data, we generated spectra from the NGC2264 best-fit parameters.
We then added  to these modeled spectra random Gaussian noise with an rms of $\sim 0.1$~K matching the median rms noise of the fitted pixels in NGC2264.
For the pixels where we did not fit emission, we used the original spectra.
The mock datacube was created at the same \kms{0.14} spectral resolution as the version used in our original analysis. 

We put the mock datacube through the same pipeline described in \refsec{sec:methods}, including the \ac{ecr} `Step 0' stage, where we fit the data using the original \kt{} parameters as initial guesses. As the mock data undergoes the same analysis pipeline as the input data, the multiple-component S/N cut of 7 is applied, as well as the global S/N cut of 3.

\subsection{Recovery by Number of Components}

\begin{figure}[htb!]
    \centering
    \includegraphics[width=\linewidth]{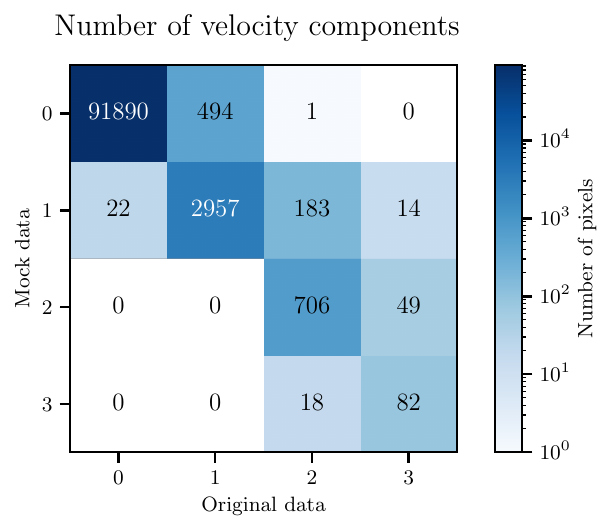}
    \caption{\label{fig:TrueMockConfusionMatrix}
    Confusion matrix comparing the number of velocity components found in the original and mock data. 
    }
\end{figure}

The final parameter cubes from the mock data contained fitted emission for 4,031 pixels, as compared to the 4,504 pixels found to have emission in the original data. 
In total 4,009 pixels were fit in both scenarios.
Generally, in the pixels where \scarplus{} identified multiple components in the input data, multiple components were also identified in the mock data. 
In this section we focus on the map-wide statistics of how many velocity components were found, in the following section we discuss on a pixel scale to what extend the parameter values are recovered. 
In \reffig{fig:TrueMockConfusionMatrix}, we compare the number of components found in the mock data using \scarplus{} to the input data. 
The comparison shows that of the 908 pixels with two-component emission in the input data (sum of third column), 706 (78\%) of those pixels were again identified as two-component in the mock version, while 18 (2\%) of those pixels were seen to have three components, 183 (20\%) were identified as one-component pixels, and a single pixel did not have detected emission in the mock data.  For the 145 three-component fits, 82 (57\%) were again identified as having three-components, 49 (34\%) were identified with two components, and 14 (10\%) were fitted with one component.

As the original data identified 4,504 pixels with emission of which only 4,009 were re-identified in the mock data, this leaves 495 pixels for which a mock spectrum was generated, but not successfully fit. 
Spatially, these pixels were primarily on the edge of the fitted regions, and were near-universally single component.
Additionally, while the original models passed the S/N cutoff of 3, these were generally dim, with a median S/N of 3.5.
In the original fitting, the median model peak of these missed pixels was 0.3~K, compared to 0.7~K for all fitted pixels. 
Given the rms noise of 0.1~K applied to the spectral models in generating the mock data, the lack of successful fitting of these pixels in the mock data is likely a result of the already weak emission being overcome by the noise.

Concerning the pixels where the mock fits found more components than were included in the input data, these were all two-component pixels for which a third was identified in the mock data.
In all of these pixels, the added third component was heavily overlapped (blended) with the pre-existing components. 

\subsection{Recovery of Parameter Values}

In addition to counting the output number of components across the map, we verified how well the fitted parameters were recovered within each pixel.
Recovery of the parameters for an individual component would indicate that such a component is well-detected in both the input and mock data. 
For the case of a spectrum in which multiple components are detected, those components may not all be detectable at the same level. 
For a single pixel, each of its input components may or may not be detected in the mock cube, and if identified, the parameters may not be consistent. 
Here we discuss the extent to which our fitting of the mock data is able to identify, with consistent parameters, individual velocity components. 

Using the 4,009 pixels with successful fits in both versions of the data, we paired up the individual components between them.
We consider a component from the mock fit to match a component from the input parameter set if it passes two criteria:
\begin{enumerate}
\item The centroid velocity must differ by no more than two channels (\kms{0.28}),
\item The velocity dispersion must be within \kms{0.14} (equal to one channel) of the input value.
\end{enumerate}

\begin{figure}[htb!]
    \centering
    \includegraphics[width=\linewidth]{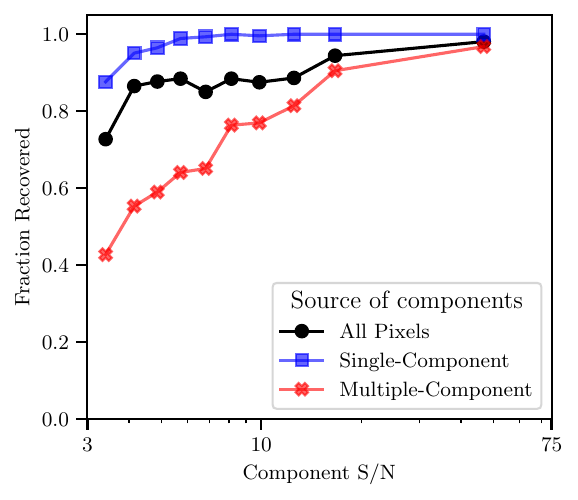}
    \caption{\label{fig:TrueComponentRecovery} Fraction of components where were recovered in the mock fits as a function of \ac{snr}. The single-component cases experience near-perfect recovery at moderate \ac{snr}.
    }
\end{figure}

Across the 4,009 pixels, there were 5,206 components in the input parameters. 
We find that 4,569 of these had a direct match in the output mock parameters, giving an overall component recovery rate of \percent{88}.
\reffig{fig:TrueComponentRecovery} breaks down the recovery rate of components by their \ac{snr}. 
We plot in black all components, but also show the recovery rates of only components from single-component pixels (blue), and only those from multi-component pixels in red.
While all groups show that recovery rate increases with \ac{snr}, the recovery rate of components from multi-component pixels is consistently lower than the rates for single-component pixels.
The data plotted in \reffig{fig:TrueComponentRecovery} are also tabulated in \reftab{tab:MockComponentRecoverySNBins}.

\begin{table*}[htb]
    \centering\caption{Component Recovery Fraction by Pixel Type
    \label{tab:MockComponentRecoverySNBins}}
\begin{tabular}{cc...}
\toprule
\multicolumn{2}{c}{\ac{snr} Bins} & \multicolumn{3}{c}{Recovery \%} \\
Percentiles & Values & \th{All Pixels}
& \th{Single Component} & \th{Multi-Component} \\
\midrule
0---10 & 3.00---3.80 & 72.7 & 87.6 & 42.8 \\
10---20 & 3.80---4.49 & 86.5 & 95.1 & 55.4 \\
20---30 & 4.49---5.24 & 87.7 & 96.5 & 59.0 \\
30---40 & 5.24---6.19 & 88.5 & 98.9 & 64.1 \\
40---50 & 6.19---7.42 & 85.0 & 99.3 & 65.1 \\
50---60 & 7.42---8.84 & 88.5 & 100.0 & 76.4 \\
60---70 & 8.84---10.90 & 87.5 & 99.6 & 77.0 \\
70---80 & 10.90---14.20 & 88.7 & 100.0 & 81.4 \\
80---90 & 14.20---19.18 & 94.4 & 100.0 & 90.5 \\
90---100 & 19.18---74.46 & 98.1 & 100.0 & 96.8 \\
\bottomrule
\end{tabular}
\end{table*}

Table \ref{tab:MockComponentRecoverySNBins} gives the breakdown of the fraction of recovered pixels in different S/N bins.  Broadly, we find that the single components are generally well recovered (at $> 80$\%) in the mock cube even at a S/N $\gtrsim 3$.  The multi-component spectra, however, are only well fit ($> 75$ \%) for a S/N $\gtrsim 7$.  This distinction could be related to our S/N cuts applied in the pipeline or to challenges in disentangling multi-component spectra.  For example, in many cases of multi-component pixels where the components were not recovered, there were fewer components in the mock spectra than in the input spectra due to  two components from the input data being replaced by a blended, wider component. Additionally, because we use a blanket noise of 0.1 K for the mock cube, weaker spectra may be lost or not recovered reliably.  For example, the multi-component spectra are recovered at $> 90$\%\ at a S/N $\gtrsim 14$, suggesting that a strong detection is needed for recovery.  

However, there were a few ($< 5$ \%) of cases at higher \ac{snr} that showed differences between the input spectral fits and the mock spectral fits unrelated to blending or low S/N recovery. In these spectra, the original and mock data had kinematic parameters which do not qualify as recovered, even when the number of identified components is the same. We find that these bright spectra have different kinematic parameters often due to differences in the component ordering, where an optically thick component is identified foreground to an optically thin component, resulting in different kinematic parameters overall. As these cases affect only a small fraction of component fits, we find that the multi-component pixels with S/N $> 14$ are generally very well recovered ($> 90$\%) by \scarplus, with components well recovered down ($> 75$\%) at a S/N $\gtrsim 7$.  For the single component fits, we find that \scarplus{} recovers the spectra very well at a S/N $ \gtrsim 3$.

\subsection{Testing Sensitivity to Empirical Thresholds}\label{app:empirical-sensitivity}

As described in \refsec{sec:ReducedFitting} and further in \refsec{sec:selecting-rsr}, several measures were taken to produce reliable spectral fits such as secondary S/N requirements for multi-component pixels and the inclusion of the \ac{rsr} and subsequent decision tree in \reffig{fig:FlowchartDecisionTree} for model selection.  
As these additional steps require empirically selected values against which to compare the spectral line fits, they may bias the models.  In this section, we examine to what extent variations in these additional measure may impact our results.

\reftab{tab:MockComponentRecoverySNBins} shows the validation testing using the nominal S/N $ \geq 7$ threshold for multi-component fits. 
While we had set this threshold to avoid selecting multi-component fits for noisy spectra (beyond our blanket requirement of S/N $ \geq 3$), a threshold of S/N $ \geq 7$ may bias our fitted results for spectra where components are close to the threshold.  
Indeed, we find that we get a good recovery at S/N $ \geq 10$ even though our threshold is at a S/N $ \geq 7$. 
To determine the impact of the S/N $ \geq 7$ threshold on our results, we tested the recovery of spectra using threshold cuts of S/N $ \geq 3$, 5, and 10. 
These cuts are applied to each multi-component fit (the two-component fit from \mufasa, the three-component \scarplus{} model, and the two-component \ac{rsr} fits), and require that at least one component reaches the value. 

Unsurprisingly, we find fewer multi-component fits with a higher threshold of S/N $ \geq 10$ and more multi-component fits with the lower thresholds.
The difference is still relatively small in all cases.
We find that only 18\% of the multi-component fits are re-classified as single-component fits in the S/N $ \geq 10$ case, and that the number of multi-component spectra increases by 21\% and 11\% in the S/N $ \geq 3$ and S/N $ \geq 5$ cases, respectively. 
In all scenarios, these reclassified pixels are primarily clustered together in regions where well-detected multi-component spectra begin to blend or fade out. 
For the lowest threshold of S/N $ \geq 3$, the total number of fitted pixels increases by 3\%, primarily from additional single-component pixels. Most of the extra fitted spectra are located in patches around the edges of detectable emission. 
Overall, we find that the choice of a S/N $ \geq 7$ provides the best balance between excluding low-emission spectra or uncertain components without excluding potentially good data.  

We further tested the RSR algorithm on the mock NGC 2264 data, assuming zero-width separation potentials of 0.14 km/s (1 channel) to 0.56 km/s (4 channels), bookending our nominal value of 0.28 km/s.  
A threshold on the zero-width separation potential of two components is used during the process of determining whether an RSR model should be rejected in favour of the original (higher component) model.
We found negligible differences in the number of components found in the spectra using these different thresholds. 
Not all pixels are fit with an RSR model, and of those, \textless~8\% resulted in a different model selection with each of the adjusted thresholds.
These results show that the RSR method is relatively consistent for our adopted 0.28 km/s difference.

\begin{equation}
    \Vec{B} = 6.8m[\text{N}87^\circ\text{W}]
\end{equation}

%% file: Appendices-3_MiscellaneousMaps.tex
\section{Velocity Component Maps}\label{app:ncmaps}
Figure \ref{fig:app_ncmap} gives the component maps of all \keystone{} clouds in a similar manner as Figure \ref{fig:ncmap} for M17.  The maps are presented in alphabetical order and include the percentage of each fitted pixel that contain one, two, or three components.
\input{Figures-NCMaps_All}

%% file: Figures-NCMaps_All.tex
\begin{figure*}[htb!]
\centering
    \includegraphics[width=0.49\linewidth]{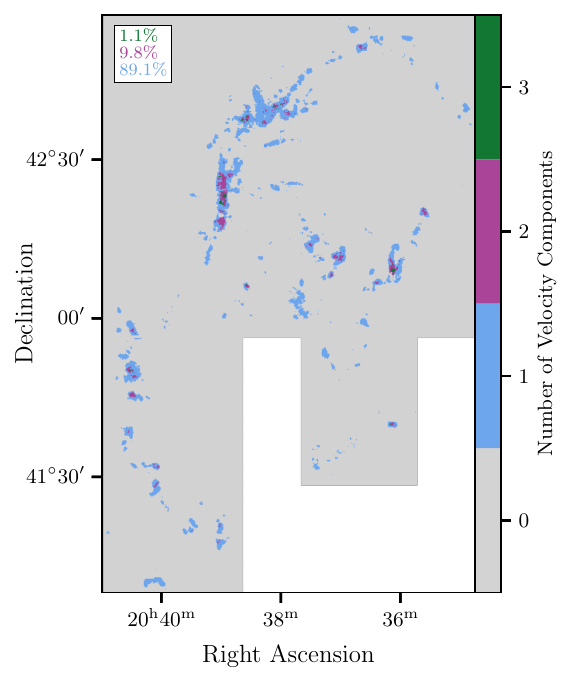}
    \includegraphics[width=0.49\linewidth]{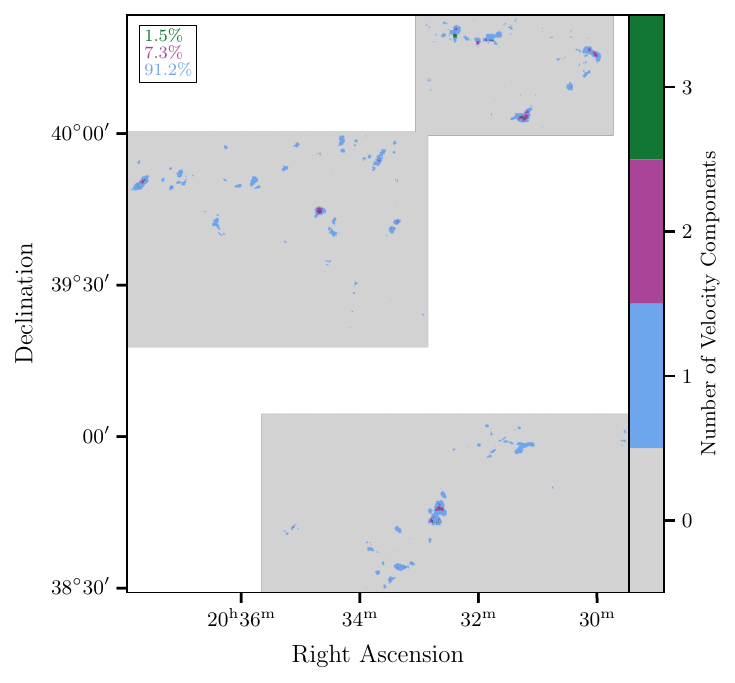} 
    \qquad
    \includegraphics[width=0.49\linewidth]{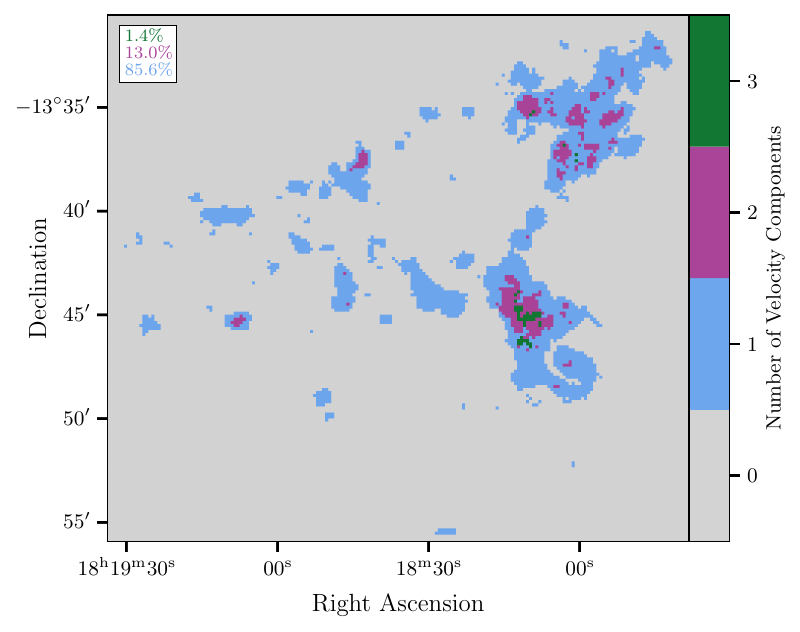} 
    \includegraphics[width=0.49\linewidth]{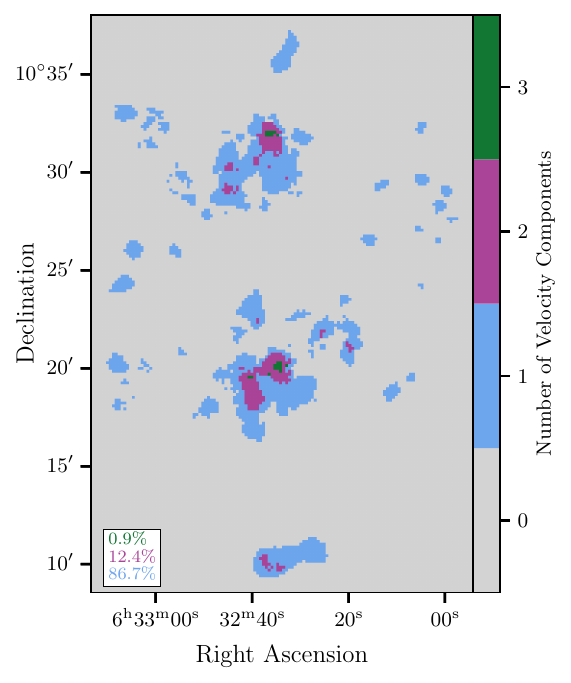}
    \caption{Component maps for the \keystone{} clouds.  Same as \reffig{fig:ncmap} for Cygnus X North (upper-left), Cygnus X South (upper-right) and M16 (lower-left), Mon R1 (lower-right)\label{fig:app_ncmap}}
\end{figure*}

\begin{figure*}[htb!]
    \centering
    \ContinuedFloat
    \includegraphics[width=0.49\linewidth]{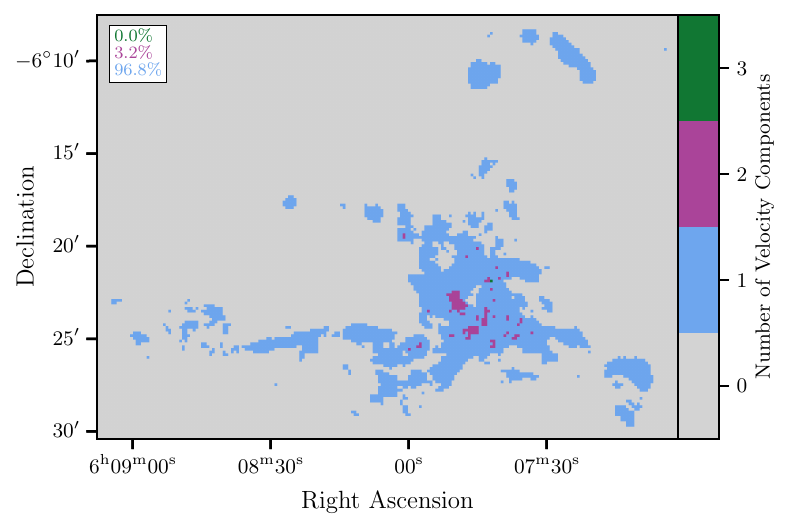}
    \includegraphics[width=0.49\linewidth]{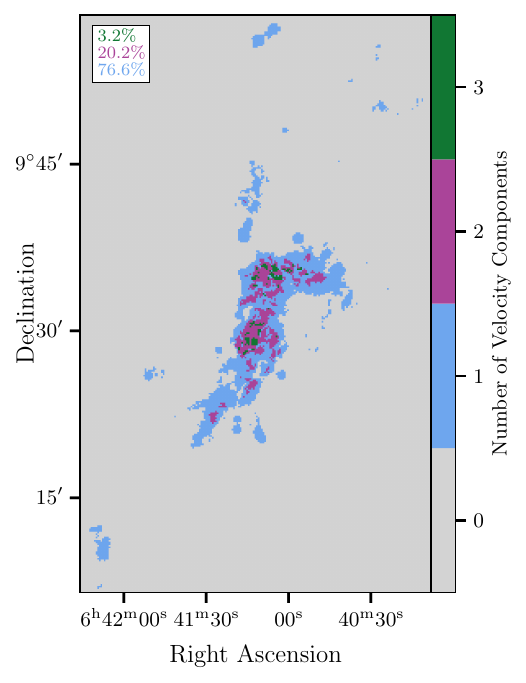}
    \qquad
    \includegraphics[width=0.49\linewidth]{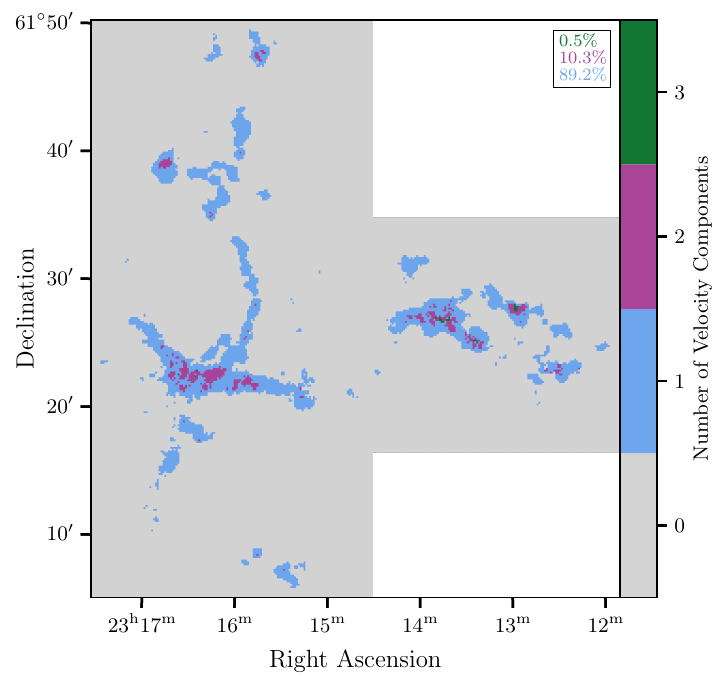}
    \includegraphics[width=0.49\linewidth]{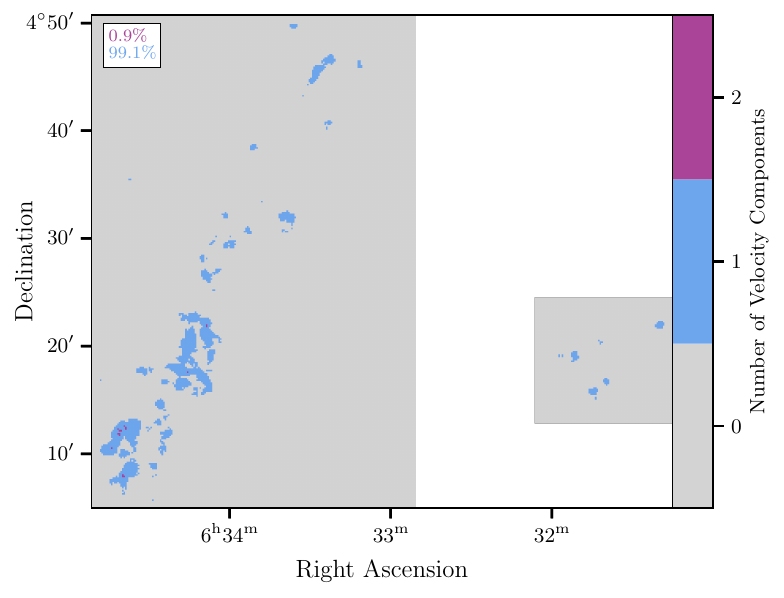}
    \caption{Continued - for Mon R2 (upper-left), NGC 2264 (upper-right), NGC 7538 (lower-left), Rosette (lower-right).}
\end{figure*}

\begin{figure*}[htb!]
    \centering
    \ContinuedFloat
    \includegraphics[width=0.49\linewidth]{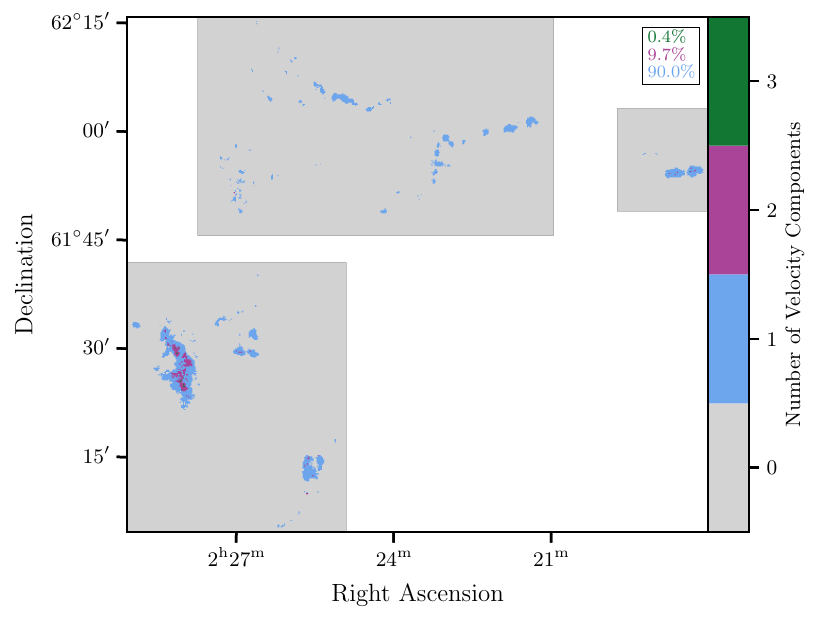}
	\includegraphics[width=0.49\linewidth]{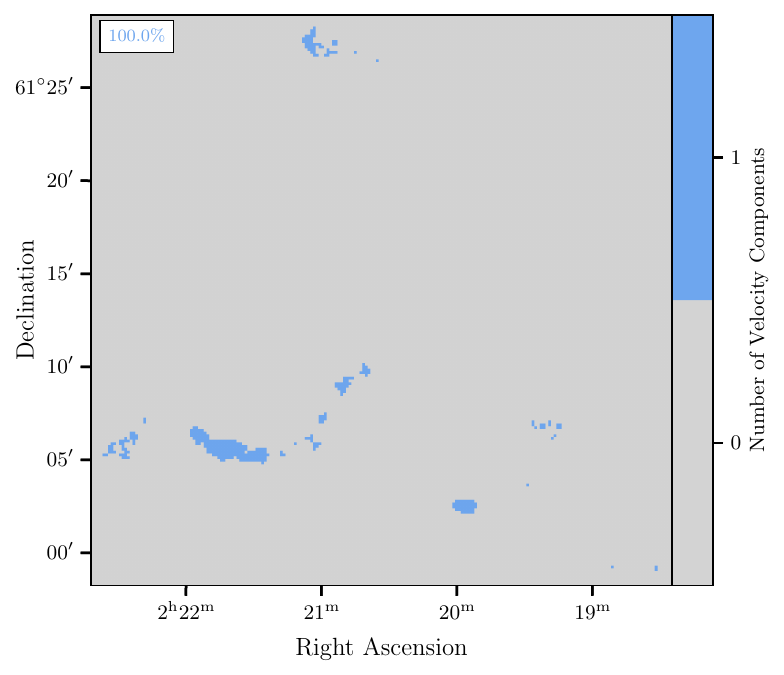}
    \qquad
    \includegraphics[width=0.65\linewidth]{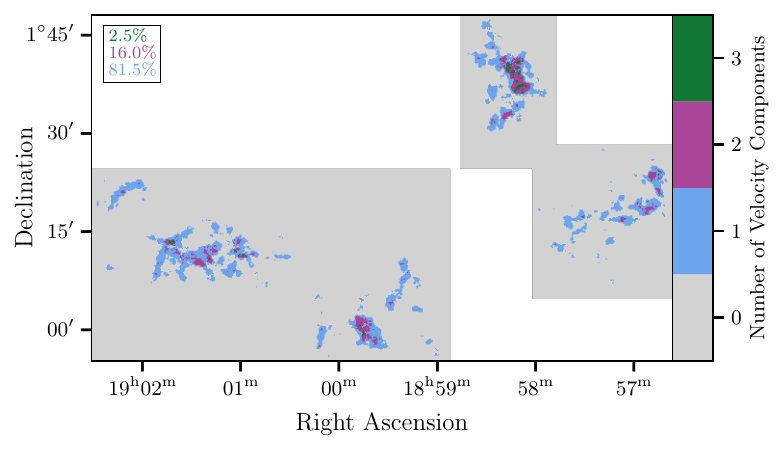}
    \caption{Continued - for W3 Main (upper-left), W3 West (upper-right), W48 (lower).}
\end{figure*}